\documentclass[a4paper, 11pt]{article}
\usepackage{geometry}

\usepackage[colorlinks=true, allcolors=blue]{hyperref}

\usepackage[english]{babel}
\usepackage{tabularx}
\usepackage{dsfont} 
\newcommand{\mathbbm}[1]{\mathds{#1}}
\usepackage{algorithm}
\usepackage{algpseudocode}
\usepackage{import}

\usepackage{amsmath, amssymb, stmaryrd, bbold, braket, amsfonts}

\usepackage{graphicx}
\usepackage{subcaption}

\usepackage{pgfplots}
\pgfplotsset{compat=1.18}
\usepackage{graphicx}
\usepackage{pgfplots}
\usepackage{subcaption}
\pgfplotsset{compat=1.18}

\title{Particle Filtering for Non-Deterministic Electrocardiographic Imaging}

\author {Emma Lagracie$^{1, 2}$, Luc de Montella$^{2}$\\
\ \\ 
$^1$ Univ. Bordeaux, IMB, UMR 5251, IHU Liryc, F-33400 Talence, France\\
$^2$ Inria Bordeaux, F-33400 Talence, France}

\begin{document}
\setlength{\emergencystretch}{3em}

\newcommand{\myplotBL}{
  \begin{tikzpicture}
    \begin{axis}[
      xlabel={Time (ms)},
      ylabel={Probability},
      xmin=0,
      xmax=120,
      grid=major,
      width=\linewidth,
      height=0.6\linewidth,
      tick label style={font=\footnotesize},
      label style={font=\footnotesize},
    ]
      \addplot[color=blue, thick] table[col sep=comma, x=time, y=vrai] {data/bloc_distribution_41.csv};
      \addplot[color=red, thick] table[col sep=comma, x=time, y=faux] {data/bloc_distribution_41.csv};
      \addplot[color=green!60!black, thick] table[col sep=comma, x=time, y=no_bloc] {data/bloc_distribution_41.csv};
    \end{axis}
  \end{tikzpicture}
}

\newcommand{\myplotTL}{
  \begin{tikzpicture}
    \begin{axis}[
      ylabel={Probability},
      xmin=0,
      xmax=120,
      grid=major,
      width=\linewidth,
      height=0.6\linewidth,
      tick label style={font=\footnotesize},
      label style={font=\footnotesize},
    ]
      \addplot[color=blue, thick] table[col sep=comma, x=time, y=vrai] {data/bloc_distribution_4.csv};
      \addplot[color=red, thick] table[col sep=comma, x=time, y=faux] {data/bloc_distribution_4.csv};
      \addplot[color=green!60!black, thick] table[col sep=comma, x=time, y=no_bloc] {data/bloc_distribution_4.csv};
    \end{axis}
  \end{tikzpicture}
}

\newcommand{\myplotTR}{
  \begin{tikzpicture}
    \begin{axis}[
      xmin=0,
      xmax=120,
      grid=major,
      width=\linewidth,
      height=0.6\linewidth,
      tick label style={font=\footnotesize},
      label style={font=\footnotesize},
    ]
      \addplot[color=blue, thick] table[col sep=comma, x=time, y=vrai] {data/bloc_distribution_5.csv};
      \addplot[color=red, thick] table[col sep=comma, x=time, y=faux] {data/bloc_distribution_5.csv};
      \addplot[color=green!60!black, thick] table[col sep=comma, x=time, y=no_bloc] {data/bloc_distribution_5.csv};
    \end{axis}
  \end{tikzpicture}
}

\newcommand{\myplotBR}{
  \begin{tikzpicture}
    \begin{axis}[
      xlabel={Time (ms)},
      xmin=0,
      xmax=120,
      grid=major,
      width=\linewidth,
      height=0.6\linewidth,
      tick label style={font=\footnotesize},
      label style={font=\footnotesize},
    ]
      \addplot[color=blue, thick] table[col sep=comma, x=time, y=vrai] {data/bloc_distribution_51.csv};
      \addplot[color=red, thick] table[col sep=comma, x=time, y=faux] {data/bloc_distribution_51.csv};
      \addplot[color=green!60!black, thick] table[col sep=comma, x=time, y=no_bloc] {data/bloc_distribution_51.csv};
    \end{axis}
  \end{tikzpicture}
}

\maketitle
\begin{abstract}
Electrocardiographic imaging (ECGI) aims to non-invasively reconstruct activation maps of the heart from temporal body surface potentials. While most existing approaches rely on inverse and optimization techniques that may yield satisfactory reconstructions, they typically provide a single deterministic solution, overlooking the inherent uncertainty of the problem stemming from severe ill-posedness, limited knowledge of biophysical properties, and unavoidable measurement noise. The Bayesian framework, which naturally incorporates uncertainty while accounting for temporal correlations, offers a principled way to address these limitations.

In this work, we propose a low-dimensional representation of the cardiac activation sequence to enable the use of particle filtering. Unlike Kalman-based approaches, this Bayesian estimation method does not rely on predefined assumptions regarding the shape of the posterior distribution. This enables the production of not only activation maps but also probabilistic maps indicating the likelihood of activation at each point on the heart over time, as well as pseudo-probability maps reflecting the likelihood of a point being part of an earliest activation site. Additionally, we introduce a method to estimate the probability of the presence of conduction lines of block (LOBs) on the heart surface. Combined with classical reconstruction techniques, this could help discriminate artificial from true LOBs in activation maps. 

Results using simulated data show that, in favorable settings, the method accurately recovers both the number and location of earliest activation sites and effectively discriminates true from false LOBs. Under strong model mismatch, the loss of accuracy is balanced by increased uncertainty, correctly reflecting lower confidence levels.
\end{abstract}


\section{Introduction}
Preceding any contraction, the heart cells undergo an electrical stimulation, called the action potential, which spreads throughout the entire muscle to synchronize the contraction of the heart. If abnormal, this \textsl{activation sequence} can trigger malfunctions in the cardiac rhythm. Thus, the electrical activity of the heart provides valuable insight for detecting and identifying cardiac rhythm pathologies.

Electrocardiographic imaging, also called ECGi, aims to non-invasively recover information on the electrical activity of the heart. Using mathematical models of the electrical activity of the heart and torso (source models), ECGi involves solving an inverse problem to reconstruct the bioelectrical sources in the heart (such as transmembrane voltage or extracellular potential) responsible for electrical potential measurements on the torso surface (BSPMs) \cite{pullan2010inverse, bear2015introduction, cluitmans2015noninvasive}. In particular, the goal is often to reconstruct the activation sequence of the heart, a quantity of great clinical interest \cite{schuler2021reducing}. 

However, the inverse problem of ECGi is highly ill-posed in the sense of Hadamard, meaning that at least one of the following conditions fails: existence of a solution, uniqueness of the solution or continuous dependence on the data. In practice, the ill-posedness of the problem is characterized by a very high sensitivity to small changes in the data, making the ECGi problem a very difficult challenge \cite{belgacem2007cauchy}. Moreover, in the clinical setting, source models inevitably involve approximations, such as the omission of heart and torso motion over time, or of the presence of electrical conductivity inhomogeneities caused by organs (lungs, bones, etc.) within the torso. These modeling inaccuracies further increase the complexity of the problem.

Most often, the ECGi problem is formulated as a static Cauchy problem for the Laplace equation in the torso, combined with a Tikhonov regularization term to address its ill-posedness \cite{bear2015introduction, cluitmans2015noninvasive, belgacem2007cauchy, tikhonov1977solutions}. This allows recovery of the epicardial potential distribution at multiple time instants, from which an activation map can be built \cite{schuler2021reducing}. The latter indicates, for each point of the heart, its moment of electrical activation. However, due to the inherent difficulty of the problem, the reconstructions are often not sufficiently accurate. Moreover, in a context involving very partial measurement data, limited knowledge of biophysical electrical properties and strong ill-conditioning, relying on deterministic source models and offering a single deterministic reconstruction does not seem entirely relevant. Uncertainty could in principle be accounted for within deterministic optimization frameworks by explicitly parameterizing variability \cite{opper2009variational}, however such formulations are rarely adopted in practice.

Bayesian filtering \cite{sarkka2023bayesian} provides a powerful alternative to tackle the inverse problem of cardiac electrophysiology by rigorously integrating both modeling uncertainties and measurement noise. Furthermore, rather than seeking a single deterministic solution to the ECGi problem, it estimates the full probability distribution of the quantity of interest, called \textsl{state} (e.g., extracellular potential or transmembrane voltage), conditioned on the observations (e.g., BSPMs data). This not only avoids forcing a choice among multiple plausible states but also provides tools for confidence assessment. In addition, it naturally introduces state dynamics, which may improve the conditioning of the inverse problem.

Assuming linear Gaussian state dynamics and state-to-observation relationship, the filtering problem can be solved exactly using the Kalman filter. As in related fields for decades~\cite{585768, 568913, schmidt1999bayesian}, the Kalman filter has also been applied to the ECGi problem, see \cite{aydin2011kalman, goussard1993time, schulze2013kalman} and the comprehensive review \cite{dogrusoz2019statistical}. To handle more complex non-linear state dynamics or state-to-observation models, sub-optimal Kalman-based methods, such as the extended Kalman filter and the unscented Kalman filter \cite{wan2000unscented}, have been used in electrocardiographic imaging \cite{wang2009physiological, liu2010noninvasive}. 

However, while extensions to more general distributions have been proposed, for example using non-linear transformation of a Gaussian variable \cite{Fletcher2023Lognormal, Collin2022Estimation}, most Kalman-based methods still rely on the assumption of a Gaussian posterior state distribution, which can be overly restrictive \cite{wang2009physiological}. In addition, most studies focus only on the maximum a posteriori estimate, thereby identifying the most probable state while discarding valuable information from the full probability density. Although some studies examine how well the filter’s covariance predicts the actual reconstruction error \cite{serinagaoglu2005bayesian}, restricting the solution to a Gaussian distribution limits error analysis. Finally, although Gaussian mixture representations have been proposed \cite{Li2016Gaussianmixture, Ahmed2024GaussiamSum}, classical Kalman-based methods remain limited in their ability to represent multimodal posterior distributions, potentially affecting estimation accuracy \cite{4527201}.

Particle filtering, also known as sequential Monte Carlo methods, provides an alternative to Kalman-based filters without assuming a specific shape for the target distribution \cite{liu1998sequential}. For instance, it is well suited to handle complex multimodal distributions and mixed discrete–continuous state spaces \cite{ristic_beyond_2004}. These methods have been successfully applied in many fields \cite{haupt2007genetic, lopes2011particle, iglesis2024target}, and are supported by strong theoretical results \cite{DELMORAL1999429, chopin2020introduction, caffarel2025mathematical}. However, they are computationally intensive, which has so far prevented their use in electrocardiographic imaging for reconstructing activation sequences. For instance, the authors in \cite{wang2009physiological} note that particle methods were not feasible for their model due to the high dimensionality of the state (the transmembrane voltage across the heart volume).

To overcome this issue, we introduce a parameterized representation of the cardiac activation sequence, with the goal of drastically reducing the dimensionality of the state. The main parameters used to describe the transmembrane voltage distribution over the whole transmural volume are a set of $l$ points on the heart and $l$ associated positive real numbers, where $l$ is typically less than four. These points can be interpreted as activation centers, or earliest activation sites, and each real number as the radius of a circle (with respect to a geodesic norm adapted to the heart's geometry and conductivity) within which cardiac cells are considered activated. From these parameters and a predefined activation shape, we reconstruct the transmembrane voltage. As a result, the dimensionality of the state (the $l$ centers and $l$ radii) is no longer a barrier to applying particle filtering.

Predefined templates for the transmembrane voltage have already been extensively used in ECGi, as they constitute an efficient way to constrain the space of solutions of the inverse problem \cite{van2009non, ravon2017parameter, fehrenbach2023source, franzone1990mathematical, grandits2021geasi}. In our setting, it also simplifies the choice of the state evolution model, usually a difficult but essential step in making Bayesian estimation methods effective \cite{aydin2011kalman, erenler2019ml, dogrusoz2020use, wang2009physiological}. In this work, we deliberately choose a very simple one to emphasize that our method can be effective even without a complex dynamic model or strong prior knowledge.

The chosen parameterization makes the state–to-observation relationship highly nonlinear, further motivating the use of particle filtering. More importantly, this particle filtering method enables the computation of clinically relevant quantities beyond the heart’s activation map and supports more detailed confidence assessments. For instance, it allows the calculation of the conditional probability, with respect to the data, for a given point on the heart to belong to an earliest activation site or to be activated at a specific time step. To address more complex scenarios, the parameterization can be extended by introducing additional dimensions, such as incorporating a choice among multiple geodesic distance metrics. This extension can be used to assess confidence in the presence of a true conduction block at a specific location. To our knowledge, this is the first approach enabling such an analysis.

The remainder of the paper is organized as follows. In Section~\ref{sec:background}, we recall the formulation of the electrocardiographic imaging problem, as well as the principles of Bayesian estimation and particle filtering. Section~\ref{sec:method_description} is devoted to describing our low-dimensional representation of the cardiac activation, along with the equations governing its temporal evolution and its relationship to the observed data. The numerical results supporting our method are presented in Section~\ref{sec:numerical_results}, with estimation tools related to the activation sequence detailed in Subsection~\ref{subsec:estimation_activation_sequence}, and modifications aimed at distinguishing artificial from true block lines discussed in Subsection~\ref{subsec:distinguish_block}. 

\section{Background}\label{sec:background}
\subsection{Cardiac Electrophysiology and the ECGI Problem}
In mathematical cardiac electrophysiology, the bidomain equations \cite{lines2003modeling, tung1978bi} constitute the reference model for computing the dynamics of the electrical potentials and voltages in the heart and torso domains. We denote $\Omega_H$ and $\Omega_T$ the heart and torso domains respectively, $\Omega = \Omega_H \cup \Omega_T$ the whole domain and $\Gamma_T$ the torso surface. 
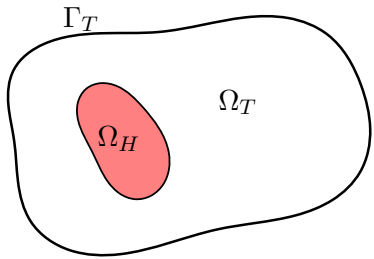
\begin{figure}[h!]
    \centering
	\def\svgwidth{0.3\textwidth}
	\centering
\begingroup%
  \makeatletter%
  \providecommand\color[2][]{%
    \errmessage{(Inkscape) Color is used for the text in Inkscape, but the package 'color.sty' is not loaded}%
    \renewcommand\color[2][]{}%
  }%
  \providecommand\transparent[1]{%
    \errmessage{(Inkscape) Transparency is used (non-zero) for the text in Inkscape, but the package 'transparent.sty' is not loaded}%
    \renewcommand\transparent[1]{}%
  }%
  \providecommand\rotatebox[2]{#2}%
  \newcommand*\fsize{\dimexpr\f@size pt\relax}%
  \newcommand*\lineheight[1]{\fontsize{\fsize}{#1\fsize}\selectfont}%
  \ifx\svgwidth\undefined%
    \setlength{\unitlength}{471.94354183bp}%
    \ifx\svgscale\undefined%
      \relax%
    \else%
      \setlength{\unitlength}{\unitlength * \real{\svgscale}}%
    \fi%
  \else%
    \setlength{\unitlength}{\svgwidth}%
  \fi%
  \global\let\svgwidth\undefined%
  \global\let\svgscale\undefined%
  \makeatother%
  \begin{picture}(1,0.68246656)%
    \lineheight{1}%
    \setlength\tabcolsep{0pt}%
    \put(0,0){\includegraphics[width=\unitlength,page=1]{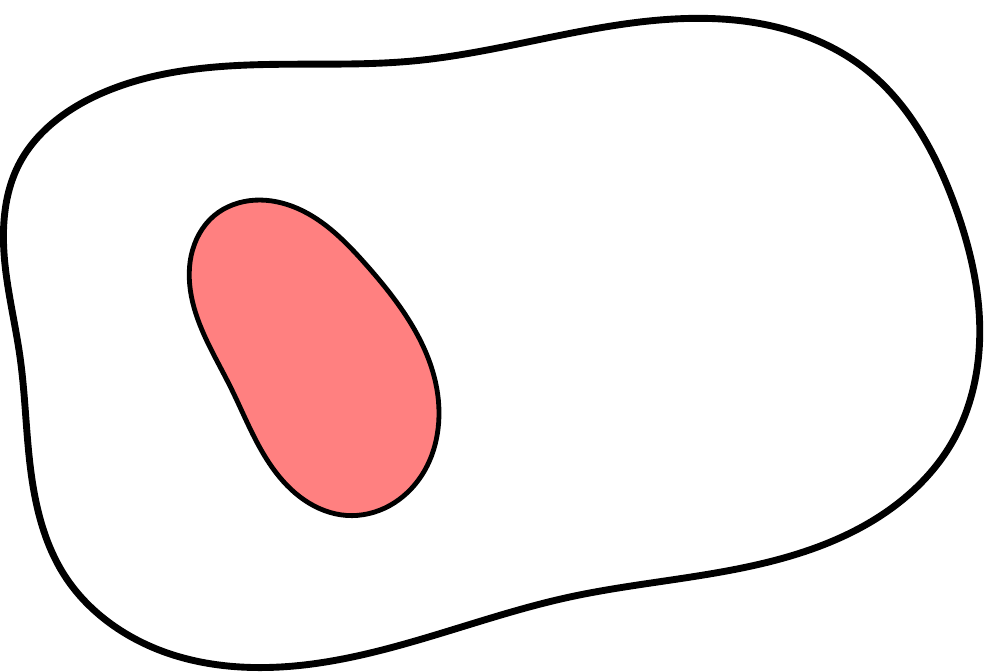}}%
    \put(0.24872773,0.30476723){\color[rgb]{0,0,0}\makebox(0,0)[lt]{\lineheight{1.25}\smash{\begin{tabular}[t]{l}$\Omega_H$\end{tabular}}}}%
    \put(0.58022486,0.40536875){\color[rgb]{0,0,0}\makebox(0,0)[lt]{\lineheight{1.25}\smash{\begin{tabular}[t]{l}$\Omega_T$\end{tabular}}}}%
    \put(0.15481671,0.63417075){\color[rgb]{0,0,0}\makebox(0,0)[lt]{\lineheight{1.25}\smash{\begin{tabular}[t]{l}$\Gamma_T$\end{tabular}}}}%
  \end{picture}%
\endgroup%

    \caption{Schematic representation of the domain $\Omega = \Omega_H \cup \Omega_T$.}
    \label{schema bidomain}
\end{figure}
The transmembrane voltage~$v\in H^1(\Omega_H)$ and the extracellular and extracardiac potential~$u \in H^1(\Omega)$ satisfy the following equations, written in a non-dimensional form: for all $t \in [0, T]$,
\begin{equation}
    \left \{ 
    \begin{aligned}
        &\text{div}(\sigma_i \nabla (u + v)) = \partial_t v + f(v, t)& \text{in } \Omega_H,\\
        &\text{div}((\sigma_i+\sigma_e) \nabla u) = -\text{div}(\sigma_i \nabla v)& \text{in } \Omega_H,\\
        & \text{div}(\sigma_T \nabla u) = 0 & \text{in } \Omega_T,\\
        &\sigma_i \nabla (u+v) \cdot n = 0 & \text{on } \partial \Omega_H,\\
        & \sigma_T \nabla u \cdot n  = 0 & \text{on } \Gamma_T,
    \end{aligned}
    \right.
    \label{bidomain}
\end{equation}where $\sigma_{i, e, T}$ are scaled conductivity tensors of the intracellular, extracellular and torso domains respectively, and $f$ is a scaled ionic current, that also depends on a set of coupled ODEs. Due to the reaction-diffusion equation, during the activation process of the heart, an action potential spreads as a wave accross the whole myocardium, offering a very characteristic time course to the transmembrane voltage at each point of the heart. Thus, its shape, contrary to the one of the extracellular potential, can be parametrized with acceptable accuracy using a predefined template.

Based on part of the bidomain equations, and a priori knowledge of cardiac electrophysiology, the inverse ECGi problem consists in recovering the transmembrane voltage $v$ or the extracelular potential $u$ distribution in the heart volume or on the heart surface (epicardium or epi-endocardial surface), to further retrieve an activation map \cite{pullan2010inverse, bear2015introduction, li2024solving, cluitmans2015noninvasive}. The most common deterministic approaches seek to reconstruct the epicardial extracellular potential distribution by solving a Cauchy problem for the Laplace equation \cite{bear2015introduction, van2023basis} (static formulation) or use a predefined template shape for the transmembrane voltage on the epicardial surface to formulate an activation-time based non-linear inverse problem \cite{van2009non, van2023basis} (intrinsically temporal formulation). 

For a clinical use, the quantity of interest is the activation map, in which some features may indicate a pathology. For instance, it is usual to identify the earliest activation sites of the heart, and so the origin of premature ventricular contractions (PVC), or to look at low conduction regions and lines of block in the map, which may trigger ventricular fibrillation \cite{duchateau2019performance}. However, we usually lack the means to assess the level of confidence in the reconstructions of the activation maps and its clinical outcome.

\subsection{Bayesian Estimation}
Bayesian estimation is a method for computing the probability distribution of a state, i.e. the quantity of interest such as the extracellular potential or the transmembrane voltage, using observations of that state (e.g. BSPMs data). It consists in sequentially incorporating data through two key models: a \textsl{state model} that describes the stochastic evolution of the state between consecutive measurements, and a \textsl{measurement equation} that characterizes the probabilistic relationship between the state and the observed data.
 
 The state is assumed to be a Markov chain $(X_k)_{k \in \mathbb{N}}$ on $\mathbb{R}^d$, for some $d \in \mathbb{N}^*$. It evolves according to the \textsl{state equation}:
\begin{equation}
X_{k+1} = f_k(X_k, v_k), \label{eq:state_equation}
\end{equation}
where $(v_k)_{k \in \mathbb{N}}$ is a family of independent and identically distributed (i.i.d.) random variables on $\mathbb{R}^d$. These variables model both the random evolution of $X_k$ and uncertainty in the model function $f_k$. 

The observation $Y_k \in \mathbb{R}^q$ at time $k$ is linked to the state by the \textsl{measurement equation}:
\begin{equation}
Y_k = h_k(X_k, w_k), \label{eq:measurement_equation}
\end{equation}
where $(w_k)_{k \in \mathbb{N}}$ is a family of i.i.d. random variables on $\mathbb{R}^q$. These account for both the measurement noise and the uncertainty in the measurement function $h_k$.

Bayesian estimation aims to compute the conditional probability density, denoted $p(X_k \mid Y_{1:k})$, of the state $X_k$ given the observations $Y_{1:k} := (Y_1, \dots, Y_k)$ up to time $k$. This computation is done iteratively with each new observation, via two steps.

Assuming $p(X_k \mid Y_{1:k})$ is known, the \textsl{prediction step} forecasts the state density at the next time step using the transition kernel known through the \textsl{state equation}~\eqref{eq:state_equation}:
\begin{equation}
p(X_{k+1} \mid Y_{1:k}) = \int_{\mathbb{R}^d} p(X_{k+1} \mid X_k) \, p(X_k \mid Y_{1:k}) \, dX_k. \label{eq:prediction_bayesian}
\end{equation}
In the \textsl{correction step}, this prediction is then refined using the new observation and the likelihood known through the \textsl{measurement equation}~\eqref{eq:measurement_equation}:
\begin{equation}
p(X_{k+1} \mid Y_{1:k+1}) \propto p(Y_{k+1} \mid X_{k+1}) \, p(X_{k+1} \mid Y_{1:k}). \label{eq:correction_bayesian}
\end{equation}
For example, to apply the Bayesian estimation framework to electrocardiographic imaging, the observation variable \( Y_k \) is usually defined as a vector of body surface potentials, and the state variable \( X_k \) can be defined as either the vector of epicardial potentials at each node on the heart surface \cite{zhou2018localization, aydin2011kalman}, the transmembrane voltage in the myocardial volume \cite{wang2009physiological}, or an updated activation map \cite{liu2010noninvasive}.


Depending on the state one seeks to estimate, choosing a suitable state equation can be more or less complex. The temporal dynamics of the extracellular potential is less explicit than that of the transmembrane voltage, which undergoes an action potential over time. However, this choice of state equation is crucial, as it strongly influences the inverse problem reconstructions. Its deterministic component constrains the temporal evolution of the state via a prior that will have more or less importance depending on the chosen process noise. For example, in~\cite{wang2009physiological}, Wang and Lui propose to model the dynamics of the transmembrane voltage in the myocardial volume using a monodomain \cite{potse2006comparison, lines2003mathematical} reaction-diffusion model associated with a parameterized ionic model and Gaussian noise. Other works have constrained the extracellular potential on the epicardium to depend on its neighbors at the previous time step~\cite{zhou2018localization}, in order to better identify the initiation points of the heart's electrical activation. In~\cite{liu2010noninvasive}, Lui and He directly estimate the activation map of the myocardium by linking it to torso observations through a prescribed form of the transmembrane voltage and the electrostatic equilibrium part of the bidomain equations \eqref{bidomain} linking \( v \) to \( u \).

The measurement equation is usually simply composed of the state to observation transfer matrix, with an additive Gaussian noise process.

\subsection{Sequential Importance Resampling}
Particle filtering, also known as sequential Monte Carlo, is a set of methods used to approximate the Bayesian estimation equations numerically. It aims to approximate the posterior density \( p(X_k \mid Y_{1:k}) \) with a weighted sum of Dirac measures. While many sophisticated particle filtering methods exist and could prove efficient for the problem at hand~\cite{ristic2003beyond, elfring2021particle}, in this article, we restrict ourselves to the use of the Sequential Importance Resampling (SIR) algorithm~\cite{gordon1993novel}. Indeed, our aim is not to compare particle filtering techniques, but rather to propose an efficient way to apply them to ECGi problems. For this purpose, the simplicity of SIR makes it particularly well suited.

The principle of the SIR algorithm is as follows. At time \( n = 0 \), given a prior distribution density, denoted \( p(X_0) \), of the initial state \( X_0 \), we draw \( N \) independent samples \( \xi_0^i \), called \emph{particles}. Each particle is associated with a weight \( \omega_0^i \), initially set to \( \frac{1}{N} \). The initial density is then approximated by
\[
p(X_0) \approx \sum_{i=1}^N \omega_0^i \delta_{\xi_0^i}(X_0),
\]
where \( \delta_{\xi_0^i} \) denotes the Dirac measure centered at \( \xi_0^i \).

At the prediction step, the particles evolve according to the state equation~\eqref{eq:state_equation}, and the weights remain unchanged. In practice, we draw \( N \) independent realizations \( v_k^i \) of the process noise \( v_k \), and compute
\[
\xi_{k+1}^i = f_k(\xi_k^i, v_k^i), \quad \text{so that} \quad p(X_{k+1} \mid Y_{1:k}) \approx \sum_{i=1}^N \omega_k^i \delta_{\xi_{k+1}^i}(X_{k+1}).
\]

At the correction step, the particles remain unchanged, and the weights are updated according to the likelihood of the observation, following equation~\eqref{eq:correction_bayesian}:

\[
\omega_{k+1}^i = \frac{\omega_k^i \, p(Y_{k+1} \mid \xi_{k+1}^i)}{\sum\limits_{j=1}^N \omega_k^j \, p(Y_{k+1} \mid \xi_{k+1}^j)}, \quad \text{so that} \quad p(X_{k+1} \mid Y_{1:k+1}) \approx \sum_{i=1}^N \omega_{k+1}^i \delta_{\xi_{k+1}^i}(X_{k+1}).
\]

To avoid a situation where only a few particles have non-negligible weights, a phenomenon known as \emph{degeneracy}, the SIR algorithm includes a resampling step, which is triggered whenever the effective sample size
\[
N_{\text{eff}} = \frac{1}{\sum\limits_{j=1}^N (\omega_{k+1}^j)^2}
\]
falls below a predefined threshold. A new set of \( N \) particles is then drawn, with each particle selected with probability proportional to its weight. This has the effect of duplicating high-weight particles and discarding low-weight ones.

\section{Method Description}\label{sec:method_description}
\subsection{Low-Dimensional State Representation}
In this paper, we consider a state variable \( X_k \) representing the cardiac activation front. To avoid high-dimensional issues, we leverage the characteristic shape of the activation front \cite{keener1991eikonal,franzone1990mathematical} and model the electrical activation using \( l \) growing balls with respect to a geodesic measure defined on the heart mesh (volume or surface). Each ball is defined by a \textsl{center} and a time-dependent \textsl{radius}, which allows the front to be fully described with only a few parameters. From this activation front, a predefined approximation of the transmembrane voltage is computed at each time step, from which we derive the extracellular and extracardiac potentials. Since the model is based on a low-dimensional state, we can apply particle filtering methods to estimate the centers and radii of the balls, thereby reconstructing the activation map of the heart.

More formally, let us consider, at time $t_k$, \( l \) activation centers of the electrical activation of the heart, denoted by \( c^i_k \in \mathbb{R}^3 \), and referred to as \emph{centers}. Each activation point is associated with a time-dependent distance \( r^i_k \in \mathbb{R}_+ \), called the \emph{radius}. Together, the radii and centers define balls on the heart mesh (volume or surface), with respect to a geodesic distance \( d \) defined on that mesh.  This distance, which accounts for the shape of the heart, can be weighted according to the conductivity tensors defined on the heart. This, for instance, allows us to consider the fibers of the heart that give a preferential direction of propagation.

The state $X_k$ to be estimated at time $t_k$ is then:
\begin{equation} \label{eq:state_definition}
X_k = \begin{pmatrix}
c_k^1 & \cdots & c_k^l & r_k^1 & \cdots & r_k^l
\end{pmatrix}.
\end{equation}

Along with this, we consider a predefined shape of the transmembrane voltage based on a function \( V(\xi) \), where \( \xi \in \mathbb{R} \) denotes the relative distance between a point on the heart and an activation center. Here, \( V \) is expressed as a smoothed Heaviside function:
\begin{equation}
V(\xi) =
\begin{cases}
0 & \text{if } \xi < -\text{width}, \\
1 & \text{if } \xi > +\text{width}, \\
- \dfrac{\xi^3}{4\,\text{width}^3} + \dfrac{3\xi}{4\text{width}} + \dfrac{1}{2} & \text{if } |\xi| \leq \text{width}.
\end{cases}\label{shape_of_v}
\end{equation}
The reconstructed transmembrane voltage at time \(t_k\) is then defined as
\begin{equation}
    v_{\text{\tiny $X_k$}}(x) = \max_{i=1,\ldots,l} V\left(r_k^i - d(x, c_k^i)\right).
    \label{V_Xk}
\end{equation}
According to this definition, the transmembrane voltage equals \( 1 \) inside each ball defined by a center and a radius (corresponding to the activated region of the heart), and equals \( 0 \) outside (corresponding to the resting region). Between these two extreme states, a smooth transition occurs across the activation front. The resting and plateau value of the transmembrane voltage were chosen to be $0$ and $1$ in accordance to the Mitchell-Shaeffer ionic model \cite{mitchell2003two}, but can be adjusted to recover a transmembrane voltage between $-90$mV and $20$mV. It is even possible to consider these extremal values as parameters.

Note that it is not necessary to know the exact number of activation centers but only an upper bound to reconstruct the signal using our method. Indeed, if several balls overlap, the transmembrane voltage \( v \) remains in the activated state \( 1 \). Additional centers may also be associated with a null radius, and thus have no impact on the reconstruction.

From the transmembrane voltage $v$, the extracellular and extracardiac potential $u$ can be computed at each time step using the electrostatic equilibrium equations from the bidomain model \eqref{electrostic equilibrium}
\begin{equation}
    \left \{ 
    \begin{aligned}
        &\text{div}((\sigma_i+\sigma_e) \nabla u) = -\text{div}(\sigma_i \nabla v)& \text{in } \Omega_H,\\
        & \text{div}(\sigma_T \nabla u) = 0 & \text{in } \Omega_T,\\
        &\sigma_i \nabla (u+v) \cdot n + \sigma_e \nabla u \cdot n  = \sigma_T \nabla u \cdot n & \text{on } \partial \Omega_H,\\
        & \sigma_T \nabla u \cdot n  = 0 & \text{on } \Gamma_T,
    \end{aligned}
    \right.
    \label{electrostic equilibrium}
\end{equation}
in the case of a volume mesh of the myocardium, or alternatively, the Epicardial Model \cite{lagracie2025depth, lagracie2025effects} or the Equivalent Dipole Layer model \cite{yamashita1985source, geselowitz1992description, waal2023comp} in the case of a surface epicardial or epi-endocardial mesh.

\noindent\textbf{Remark 1.} \textit{The state representation given in equation~\eqref{eq:state_definition} will constitute the core of our method and will be used as such in Section~\ref{subsec:estimation_activation_sequence}. However, in order to achieve better results, or to more accurately evaluate specific aspects of the activation sequence, it is possible to extend the state. Such an extension is presented in Section~\ref{subsec:distinguish_block}, with the objective of evaluating the probability of the presence of a true conduction block line in the activation map.}

\subsection{State Dynamics and Observation Model}
In this paper, we deliberately choose to work with a very simple state equation. Our objective is to test the resilience of our method under minimal modeling assumption. We do not claim however that more elaborate state equations, whether physics-based or data-driven, could not improve the filter's performance, only that we aim to evaluate how well our method performs without relying on such additional inputs.

For the radius, the state equation reads:
\begin{align}
r_k^i &= \max\left(r_{k-1}^i + m^r_{k-1, i},\, 0\right).
\end{align}
where $m^r_{k-1, i} \sim \mathcal{N}(0, \Sigma_r^2)$. Note that with this state model, no prior is imposed on the speed of the $l$ activation fronts, and that, for instance, nothing prevent the activation front to go backward. We expect the filter to determine the propagation solely from the data, which allows us to avoid unreliable priors. Note that we choose Gaussian noise model here, but other distributions could also be considered.

Likewise, no prior is imposed on the locations of the centers, which evolve at each time step. Due to the fact that we restrict activation centers to be among heart mesh points, the overall aspect of their temporal evolution take a peculiar form. During the prediction step, we first draw $l$ random variable following an exponential distribution of parameter $\lambda$. Those random variables represent the maximum allowed displacement for each center according to the geodesic metric \( d \) defined on the heart mesh. Denoting $\mathcal{E}(\lambda)$, the exponential distribution of parameter $\lambda$, we have : 
\[d_{\max}^i \sim \mathcal{E}(\lambda)  ~~\text{and}~~\mathbf{d}_{\max} =  \left( d_{\max}^1 ~ \cdots  ~d_{\max}^l\right).
\]
The new centers \( c_k^i \), for \( j = 1,\ldots,l \), are  then drawn uniformly among the points of the heart mesh lying within the ball of radius \( d_{\max,j} \) centered on the previous center \( c_{k-1}^i \). Denoting $\mathcal{U}(S)$ the uniform distribution on some set $S$, the new state equation then reads:
\[
c_k^i \sim \mathcal{U}\big(B(c_{k-1}^i, d_{\max}^i)\big)
\]
where \( B(c_{k-1}^i, d_{\max}^i) \) denotes the ball in \( \Omega_H \), with respect to the geodesic distance $d$, centered on \( c_{k-1}^i \) with radius \( d^i_{\max} \).

In order to define our observation equation, let $O$ be the linear operator mapping $v$ to $u$ (either the electrostatic equilibrium equations \eqref{electrostic equilibrium} for a volume heart mesh, or the Equivalent Dipole Layer \cite{geselowitz1992description} or Epicardial Model \cite{lagracie2025effects} for a surface heart mesh). The observation equation of our filter is written:
\begin{equation}\label{observation_equation_ecgi}
    Y_k = O v_{\text{\tiny $X_k$}}\big|_{\Gamma_T} + w_k,
\end{equation}
where $\Gamma_T$ is the torso surface mesh, $|_{\Gamma_T}$ designates the trace operator on this surface and $w_k \sim \mathcal{N}(0, \Sigma_w)$ is a Gaussian measurement noise. Here we use Gaussian observation noise, but other noise models can be considered, provided the probability distribution of $w_k$ can be numerically approximated to compute likelihoods. In our case, the likelihood can be computed as
\begin{equation}
    p(Y_k | X_k) \propto \exp\left[{-\frac{1}{2} \left(Y_k - O v_{\text{\tiny $X_k$}}\big|_{\Gamma_T}\right)^\intercal\Sigma_w^{-1}\left(Y_k - O v_{\text{\tiny $X_k$}}\big|_{\Gamma_T}\right)}\right].
    \label{likelihood}
\end{equation}
According to our parametrization of the cardiac activation front, the observation equation is nonlinear, particularly the mapping from $X_k$ to $v_{\text{\tiny $X_k$}}$.

Since all electrical potentials are defined up to an additive constant, we also adjust in \eqref{likelihood} both the observed data $Y_k$ and the model outputs $O v_{\text{\tiny $X_k$}}\big|_{\Gamma_T}$ to have zero mean on the torso surface to allow for a meaningful comparison. 

Note that when the initial distribution provides little information, the early estimates produced by the particle filter tend to be imprecise. This is because the filter requires sufficient observational data before it can adjust the particle cloud. Since valuable information may be present in these initial moments, this limitation can be a significant drawback. However, nothing prevents us from feeding the observations to the filter in reverse chronological order, and to estimate the evolution of the potential starting from the final time. This approach is expected to allow the filter to gather information more efficiently in the initial stages. To distinguish between these two configurations, we will refer to the standard chronological ordering as the \textsl{forward} method (abbreviated as \textsl{fwd}), and the reversed ordering as the \textsl{backward} method (abbreviated as \textsl{bwd}). Without specific mention, the \textsl{backward} method is assumed throughout the paper.

The ECGi particle filtering method is summarized in Algorithm~\ref{alg:filter}.

\begin{algorithm}
\caption{ECGi SIR filter}\label{alg:filter}
    \begin{algorithmic}
        \State \textbf{Inputs:}
            \begin{itemize}
                \item The number of particles \( N \) and the number \( l \) of activation centers per particle.
                \item The sequence of $n$ body surface potential measurement vectors on the torso: \(\{Y_k\}_{k=1}^n\).

            \end{itemize}
        \State \textbf{Initialization:}
        \begin{itemize}
            \item Draw \( N \times l \) uniformly distributed activation \textit{centers} $\{ c^{i,j}_0 \mid i=1,\dots,N; \, j=1,\dots,l \}$ across the points of the mesh.
            \item Set the initial \textit{radii} $\{ r^{i,j}_0 \mid i=1,\dots,N; \, j=1,\dots,l \}$ to 1 when using the \textit{forward} method, or to 150 for the \textit{backward} method.
            \item Initialize the particles to 
            \(
                \xi_0^i = \bigl(c^{i,1}_0,\, \ldots,\, c^{i,l}_0,\, r^{i,1}_0,\, \ldots,\, r^{i,l}_0\bigr) ~~\text{for} ~~i=1,\dots,N. 
            \)
            \item Set the weights to \(\omega_0^i = \frac{1}{N}\)~~\text{for} ~~i=1,\dots,N.
        \end{itemize}
        
        \For{\(k = 1 \) \textbf{ to } \(n\)}

            \State \textbf{Prediction:}
            \begin{enumerate}
                \item Draw new \textit{radii} and \textit{centers} for all \( N \) particles \( \xi_k^i \), \( i = 1, \dots, N \):
                \begin{align*}
                    r_k^{i,j} = \max\left(r_{k-1}^{i,j} + \mathcal{N}(0, \Sigma_r^2),\, 0\right) ~\text{and}~
                    c_k^{i,j} \sim \mathcal{U}\big(B(c_{k-1}^{i,j}, \mathcal{E}(\lambda))\big).
                \end{align*}
                
            \end{enumerate}

            \State \textbf{Correction:}
            \begin{enumerate}
                \item Compute the transmembrane voltage function for each particle:
                \[
                    v_{\xi_k^i} : x \mapsto \max_{j = 1,\dots,l} V\left(r_k^{i,j} - d(x, c_k^{i,j})\right)
                \]
                
                \item Compute the unnormalized likelihood of each particle (replace \( Y_k \) with \( Y_{n-k+1} \) when using the \textit{backward} method):
                \[
                    p(Y_k \mid \xi_k^i) = \exp\left(-\frac{1}{2} \left(Y_k - O v_{\xi_k^i}\big|_{\Gamma_T}\right)^\intercal \Sigma_w^{-1} \left(Y_k - O v_{\xi_k^i}\big|_{\Gamma_T}\right)\right)
                \] 
                \item Update the weights of the particles: $\omega_k^i = \frac{\omega_{k-1}^i \, p(Y_k \mid \xi_k^i)}{\sum_{j=1}^N \omega_{k-1}^j \, p(Y_k \mid \xi_k^j)}$
            \end{enumerate}
            
            \State \textbf{Resampling:}
            \If{\( N_{\mathrm{eff}} = \frac{1}{\sum\limits_{j=1}^N (\omega_{k}^j)^2} < \frac{N}{3} \)}
                \begin{enumerate}
                    \item Resample each particle \( \xi_{k}^i \) independently from the set \( \{\xi_{k}^j\}_{j=1}^N \), with probabilities proportional to the weights \( \{\omega_{k}^j\}_{j=1}^N \).
                    \item Reset all weights to  \(\frac{1}{N} \).
                \end{enumerate}
            \EndIf
        \EndFor

    \State \textbf{Output:} The posterior density estimate at each time step: $p(X_k \mid Y_{1:k}) \approx \sum_{i=1}^N \omega_k^i \, \delta_{\xi_k^i}(X_k)$
    \end{algorithmic}
\end{algorithm}

\section{Numerical Results}\label{sec:numerical_results}
\subsection{Method and Data}

\subsubsection{Data Generation}\label{subsection:data_generation}
Data were simulated on a volume heart and torso mesh (17523 myocardial nodes, 2882 torso surface nodes and 44156 total nodes), using the bidomain equations \eqref{bidomain} coupled with the Mitchell-Shaeffer ionic model~\cite{mitchell2003two}. These equations were discretized spatially using the $P^1$-Lagrange finite elements method, and temporally with a Semi-Implicit Backward Differentiation Scheme of order 2 (SBDF2) \cite{ethier2008semi}.

\paragraph{Propagation test cases} We designed $6 \times 2$ propagation test cases for the data generation. Six primary test cases were computed with isotropic conductivity tensors $\sigma_i$ and $\sigma_e$ in the heart. Then six twin test cases were calculated with the same parameters but anisotropic conductivity conductivity tensors in the heart, using simulated ruled-based fibers from the algorithm of Bayer et al. \cite{bayer2012novel}. In the case of the isotropic (scaled) conductivities, we imposed $\sigma_i = 1 I_3$, $\sigma_e = 3 I_3$ and $\sigma_T = 2 I_3$. In the presence of the ruled-based fibers, we chose the conductivity tensors singular values according to \cite{potse2018scalable}. Denoting by ${\ell}$ the longitudinal fiber direction and ${t}$ the transverse direction, we used the following singular values: $\sigma_i^\ell = \sigma_e^\ell = 3$, $\sigma_i^t = 0.3$, $\sigma_e^t = 1.2$ and $\sigma_T = 2 I_3$. 

Out of the six primary test cases, two contain one initial stimulation site, two contain two initial stimulation sites, with a small delay between the two stimulations, and the last two cases contain one initial stimulation site but also a true conductivity line of block. These conduction line of block were computed by diminishing by 99\% the intracellular conductivity values inside a designated area. 

The test cases and their characteristics are summarized in Table \ref{tab:bidomain params}, including the true locations of the lines of block. 
\begin{table}[h!]
    \centering
       \begin{tabular}{|>{\centering\arraybackslash} m{2cm}||c|c|c|c|c|}
     \hline
    Stim & Line of & Stim  & Stim & Isotropic  & Anisotropic \\
    name & Block & localization & delay & test case name & test case name \\
     \hline 
     \hline
     Stim 1 & NA & \raisebox{-.5\height}{\includegraphics[width=0.08\linewidth]{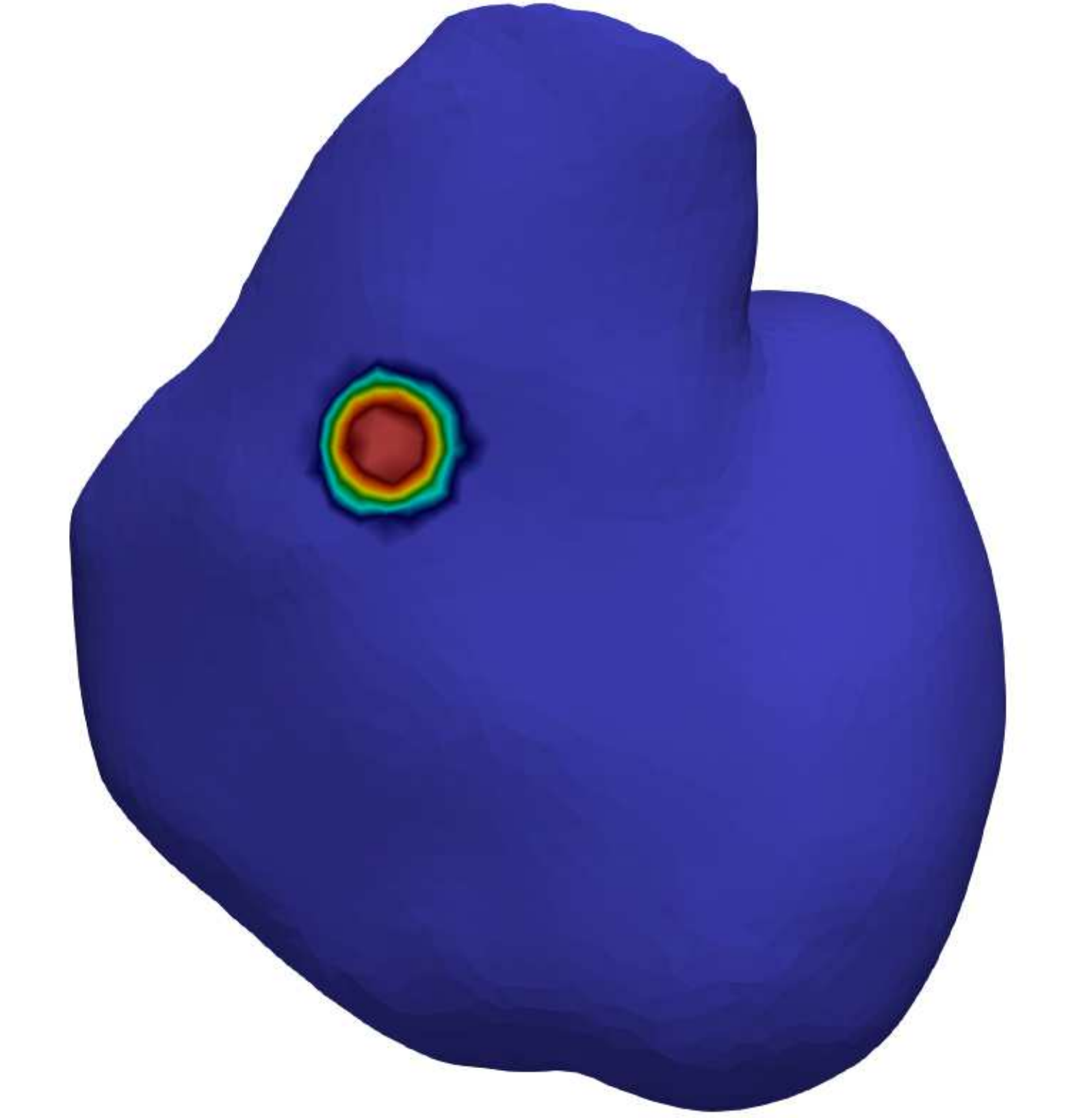}} & NA & \textsl{Stim1Iso} & \textsl{Stim1Ani}\\
      
      Stim 2 & NA & \raisebox{-.5\height}{\includegraphics[width=0.08\linewidth]{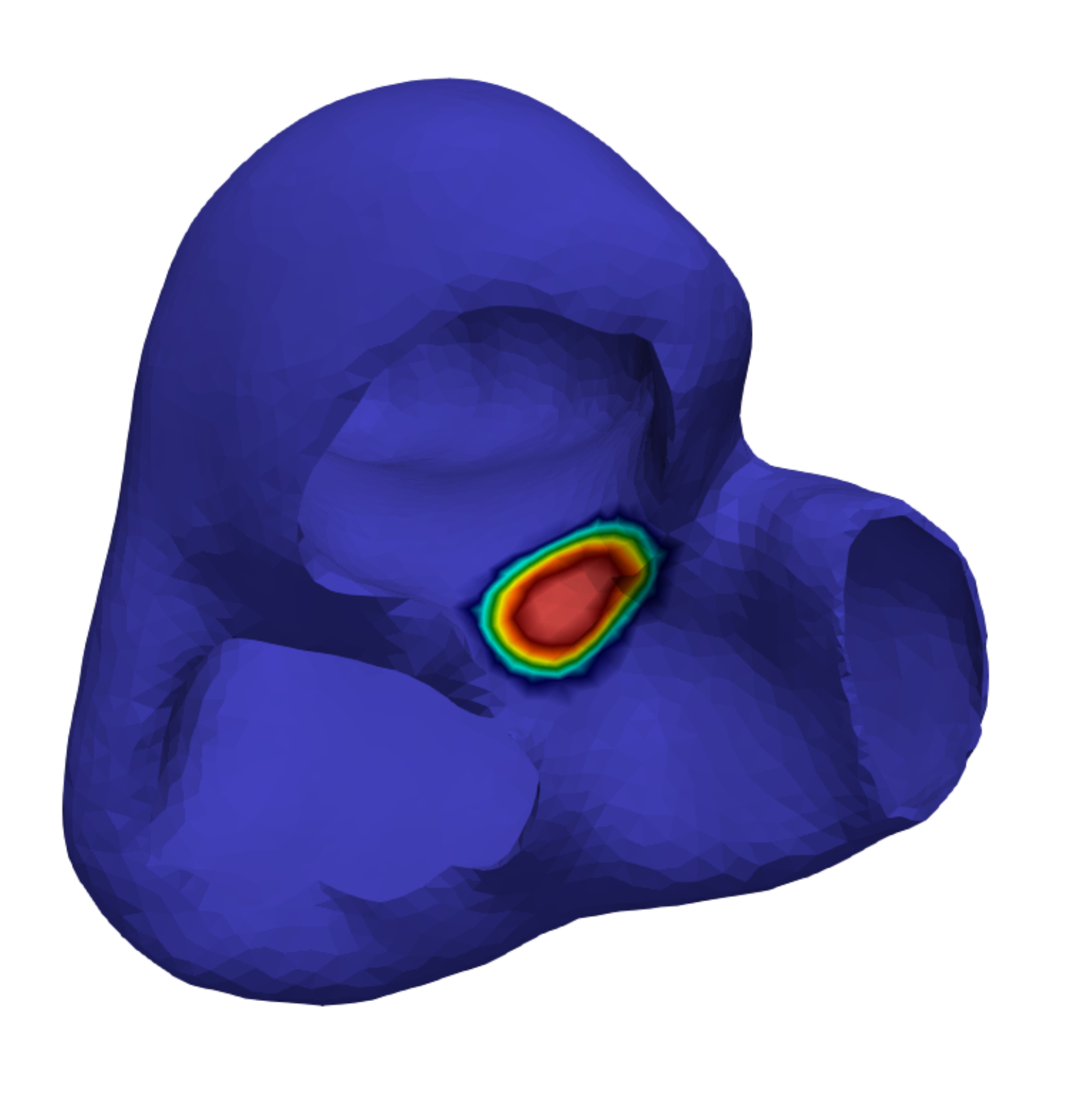}} & NA & \textsl{Stim2Iso} & \textsl{Stim2Ani}\\
      
       Delay 1 
       &NA & \raisebox{-.5\height}{\includegraphics[width=0.08\linewidth]{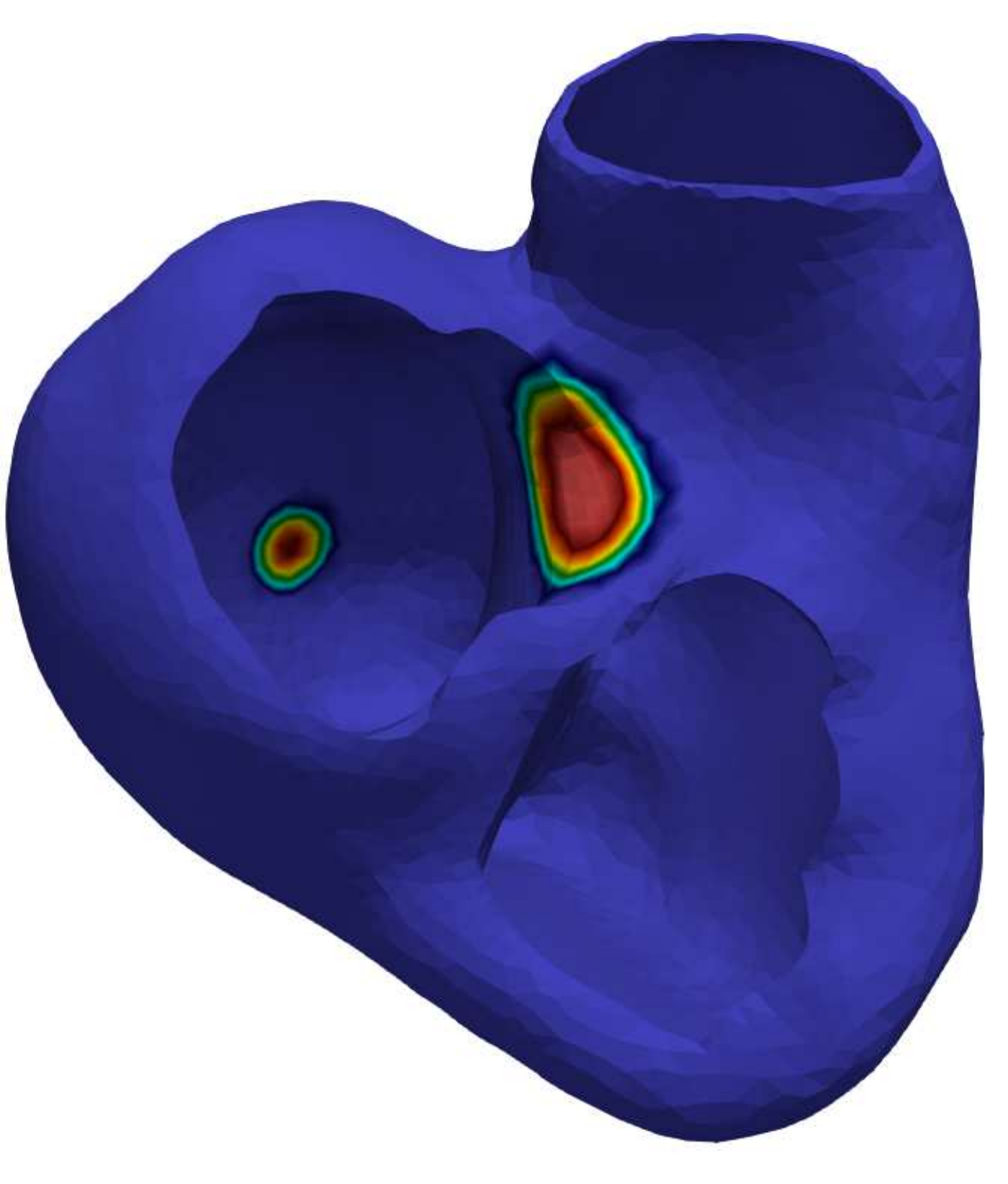}} & 15 ms & \textsl{Delay1Iso} & \textsl{Delay1Ani} \\

       Delay 2 
       & NA & \raisebox{-.5\height}{\includegraphics[width=0.08\linewidth]{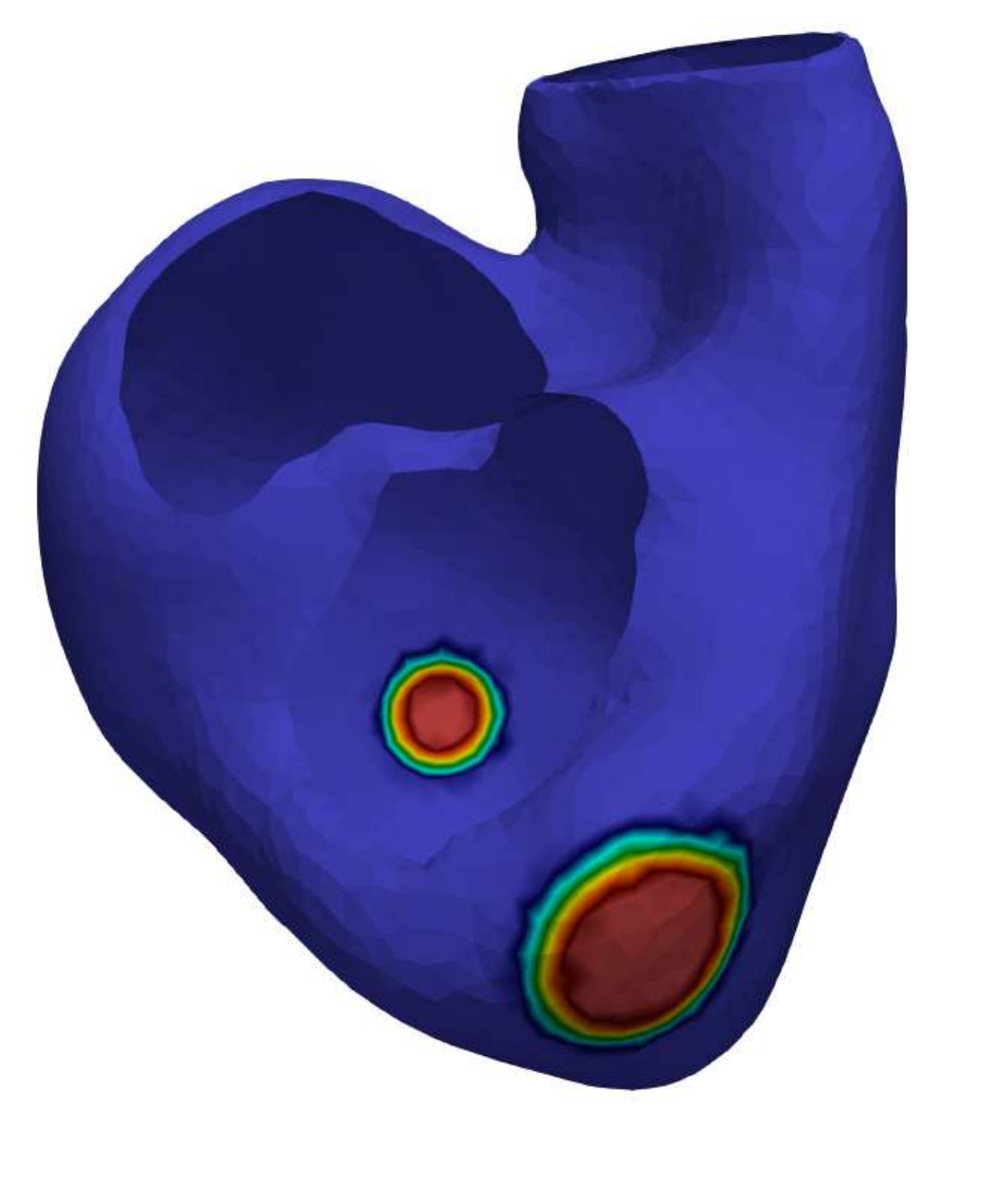}} & 20 ms & \textsl{Delay2Iso} & \textsl{Delay2Ani} \\

       Block 1 & \raisebox{-.5\height}{\includegraphics[width=0.08\linewidth]{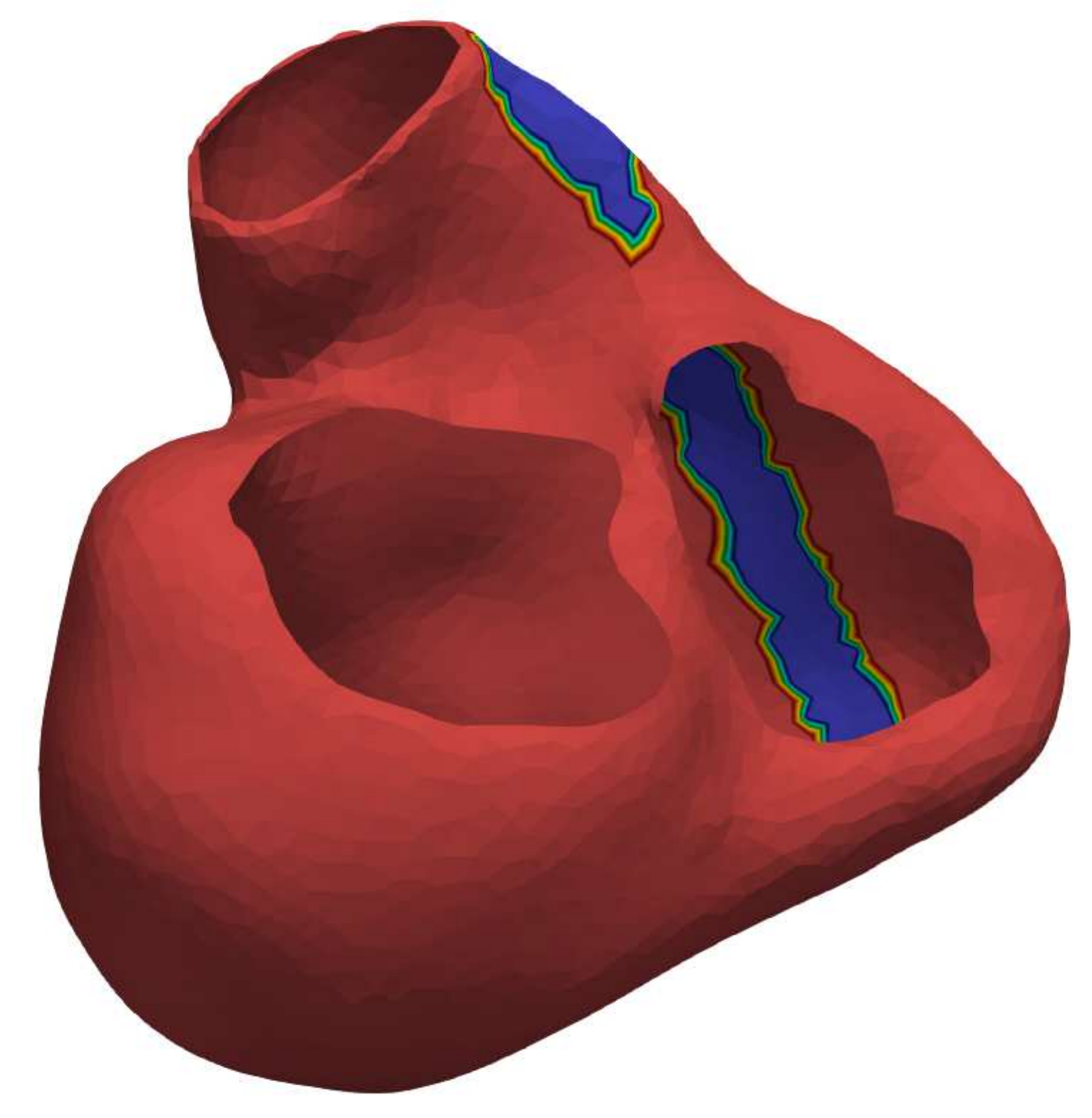}} & \raisebox{-.5\height}{\includegraphics[width=0.08\linewidth]{images/stims/stim_4_6_9.pdf}} & NA & \textsl{Block1Iso} & \textsl{Block1Ani}\\
      
      Block 2 & \raisebox{-.5\height}{\includegraphics[width=0.08\linewidth]{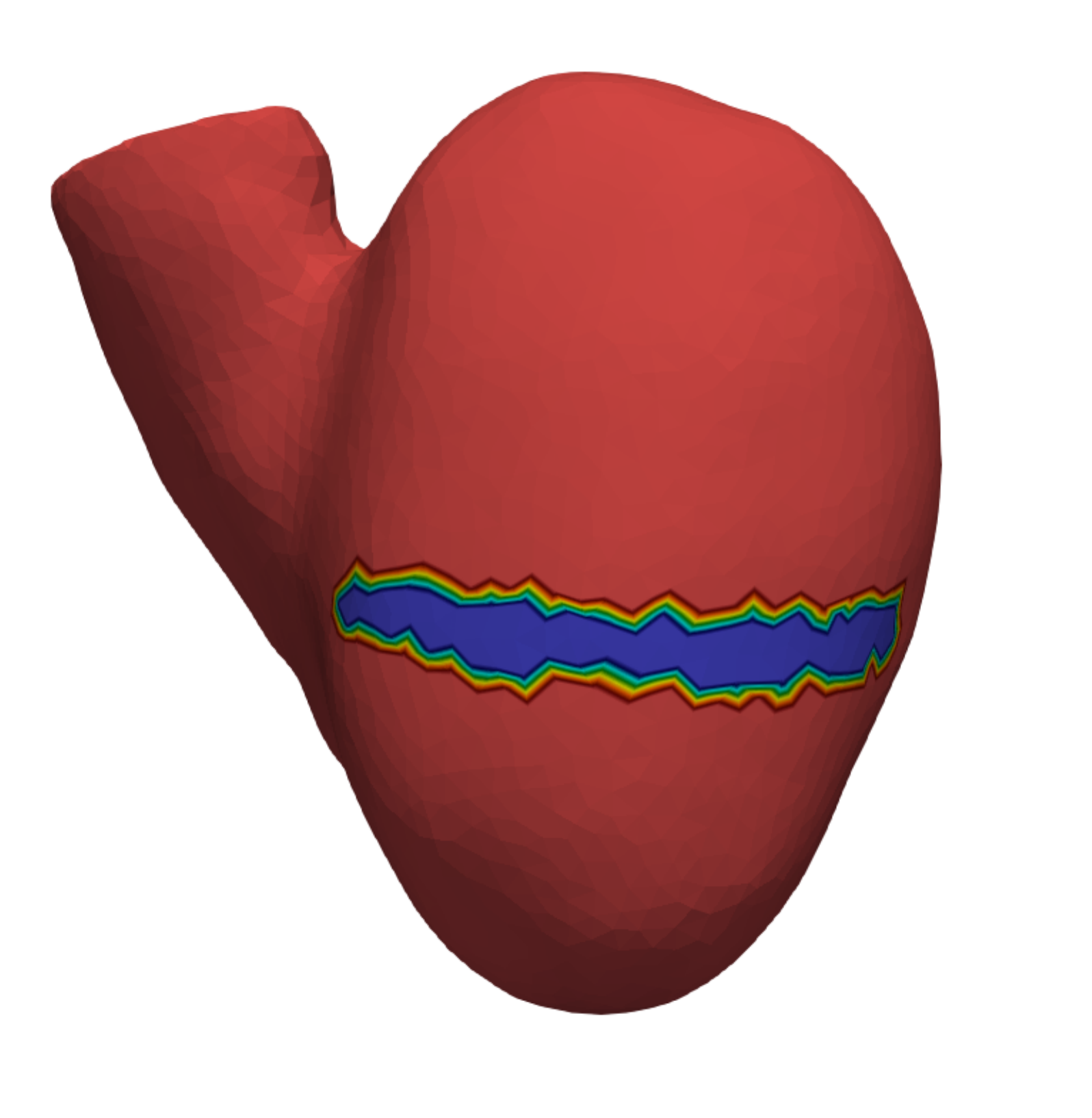}} & \raisebox{-.5\height}{\includegraphics[width=0.08\linewidth]{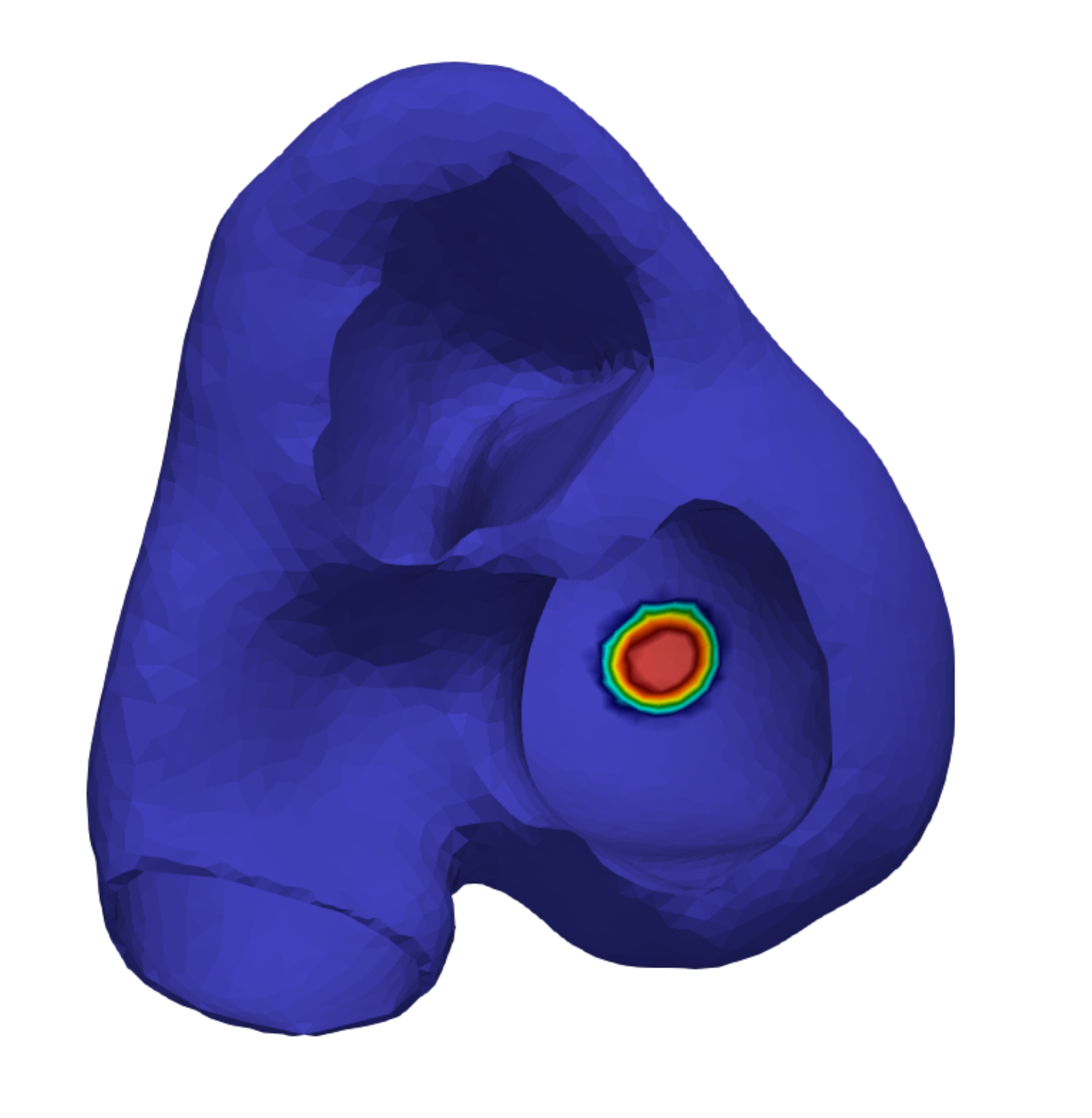}} & NA & \textsl{Block2Iso} & \textsl{Block2Ani}\\
      
      \hline
    \end{tabular}
    \caption{The $6 \times 2$ propagation test cases parameters. The first column designates the initial stimulation name. The second column presents, if applicable, the localization of the true conductivity block line. The third column gives a representation of the stimulation on the heart mesh and the fourth column indicates, if applicable, the time delay between the application of different pacings. The two last columns give the associated propagation test case name for the isotropic conductivities as well as for anisotropic conductivities.}
    \label{tab:bidomain params}
\end{table}

\paragraph{Data interpolation and noise addition} Body surface potential maps (BSPMs) were extracted from the bidomain solutions and interpolated on a second "inverse" torso surface coarser mesh (1186 nodes). Then, the interpolated torso signals were corrupted by adding white noise with a standard deviation equal to 4\% of the mean absolute amplitude of the data over time, which roughly corresponds to a signal-to-noise ratio of $30-32$~dB.

\subsubsection{Methods for the Filter}

\paragraph{Inverse mesh}
In the ECGi procedure, we used a different coarser volume heart and torso mesh of 22114 total nodes, 9174 myocardial nodes and 1186 torso surface nodes. The use of a different mesh than the one used for the data simulation allows to avoid inverse crime.

\paragraph{Transfer matrix model $O$} As we used volumetric meshes, the transfer matrix $O$ was obtained by discretizing the \textsl{isotropic} electrostatic equilibrium equations \eqref{electrostic equilibrium} with a $P^1$-Lagrange finite elements method. We imposed $\sigma_i = 1 I_3$, $\sigma_e = 3 I_3$ and $\sigma_T = 2 I_3$ for all ECGi resolutions, thus voluntarily introducing a modeling error through $O$ when applying the filter on the \textsl{Stim1Ani}, \textsl{Stim2Ani}, \textsl{Delay1Ani}, \textsl{Delay2Ani}, \textsl{Block1Ani} and \textsl{Block2Ani} data.

\paragraph{Weighted geodesic distance $d$}The geodesic distances on the heart are computed using a Fast Marching algorithm \cite{fast_marching}. To account for tissue conductivities, the weight of each vertex, initially defined as its length, is divided by a correction coefficient. Let $a$ and $b$ denote the coordinate of the endpoints of the vertex, and let $\sigma_a$ and $\sigma_b$ be the effective conductivity tensors at these points, as defined in the monodomain approximation for proportional intracellular and extracellular conductivity tensors \cite{potse2006comparison, lines2003modeling}. The correction coefficient is then given by
\begin{equation}
    \min \left( \sqrt{\frac{a^\intercal\sigma_aa}{a^\intercal a}}~,~ \sqrt{\frac{b^\intercal\sigma_bb}{b^\intercal b}} \right).
    \label{sigma_distance}
\end{equation}
Note that this approach is not a standard way to account for tissue conductivity. However, it was selected for its computational efficiency and to deliberately introduce errors into the information provided to the filtering algorithms, as it would happen in clinical settings. Alternative methods for computing geodesic distance, such as the heat method \cite{crane2017heat}, could also have been employed.

To be consistent with the chosen transfer matrix $O$, we imposed isotropic conductivities $\sigma_i = 1 I_3$ and $\sigma_e = 3 I_3$ in the geodesic distance calculation. We then expect the distance $d$ to be inaccurate to describe the activation front evolution in the anisotropic test cases \textsl{Stim1Ani}, \textsl{Stim2Ani}, \textsl{Delay1Ani}, \textsl{Delay2Ani}, \textsl{Block1Ani} and \textsl{Block2Ani}. This choice simulates a more realistic situation in which the fiber orientation in the myocardium is generally not known.

\paragraph{Front width}
The front width of the action potential for $V$ was inserted as an a priori value, defined from classical values of the front width in the transmembrane voltage solution of the bidomain model. As shown in previous work \cite{lagracie2023comparison}, some modeling errors in the front width does not generate strong errors on the output body surface potential $u$. Here, the value was set at $\text{width} = 5$mm.

\paragraph{Covariances and filter parameters} 
The number of particles is set to \( N = 1\,000 \), with each particle containing \( l = 3 \) \textsl{centers} and \textsl{radii}. Since all test cases contain at most two initial stimulation points, \( l \) is always strictly greater than the number of true stimulation points.

The initial locations of the \textsl{centers} are drawn from a uniform distribution over the points of the mesh. For the \textsl{forward} method, the radii \( r \) are initialized at \( 1\,\mathrm{mm} \). With the front width of the potential from \eqref{shape_of_v} set to \( \text{width} = 5\,\mathrm{mm} \), the resulting activated area at initialization is roughly as large as the mesh size on the heart, which corresponds to the smallest detectable value. For the \textsl{backward} method, the radii are set to \( 150\,\mathrm{mm} \), so that, given the dimensions of the domain, a large part of the heart is activated at the first time step.

Note that with this initialization method, we implicitly assume no prior knowledge of the initial particle distribution. However, if additional information is available, such as a more accurate estimate of the number and location of earliest activation sites, for example from a previous run of the filter, it is possible to adjust the initialization procedure to incorporate this knowledge into the initial particle distribution.

The standard deviation of the white noise governing the evolution of the \textsl{radii} is set to \( \Sigma_r = 10\,\mathrm{mm} \), which is deliberately large to ensure that the filter can follow the signal's evolution. The parameter governing the exponential distribution for the evolution of the \textsl{centers} is set to \( \lambda = 5\,\mathrm{mm} \), to strike a balance between allowing the centers to adjust to the true activation sites and preventing excessive displacement between time steps.

The covariance of the observation equation \eqref{observation_equation_ecgi} is set to \( \Sigma_w = (0.02)^2 I_d \), where \( I_d \) denotes the identity matrix. This value was chosen by estimating the order of magnitude of the Frobenius norm of the difference between the observed data vector and the reconstructed torso potentials from randomly sampled particles. It was then fine-tuned to allow sufficient diversity among particles while avoiding excessive resampling.

\noindent \textbf{Remark 2} (Avoiding inverse crime)\textbf{.} \textit{Several precautions have been introduced throughout the numerical experiments to avoid the so-called inverse crime. In particular, data generation and filtering are performed on different heart and torso meshes. In addition, model mismatches are intentionally introduced by using incorrect conductivity tensors in the anisotropic test cases, as well as approximate reconstructions of the geodesic distance. The filtering prior is also built with an incorrect number of earliest activation sites, and measurement noise is added to the simulated body surface potential data. Together, these choices ensure that the reported results do not rely on overly optimistic inverse settings.}

\subsection{Estimation of the Activation Sequence}\label{subsec:estimation_activation_sequence}

\paragraph{Activation maps}
We recall that the output of the filter at each time instant $t_k$ is an estimation of the posterior probability density conditioned by the data, given by
\begin{equation}
p(X_k|Y_{1:k}) \approx \sum\limits_{i = 1}^{N}{\omega^i_k \delta_{\xi_k^i
}}(X_k).
\end{equation}Using the notation \eqref{V_Xk}, an estimator of the transmembrane voltage at instant $t_k$ can then be computed as
\begin{equation}
    v(x, t_k) \approx \sum\limits_{i = 1}^{N}{\omega^i_k v_{\xi^i_k}(x)}.
\end{equation}
This estimator enables us to compute activation maps with the maximal spatio-temporal deflection method \cite{schuler2021reducing, lagracie2024assessment}.

\begin{figure}[h!]
    \centering
    \includegraphics[width=0.7\linewidth]{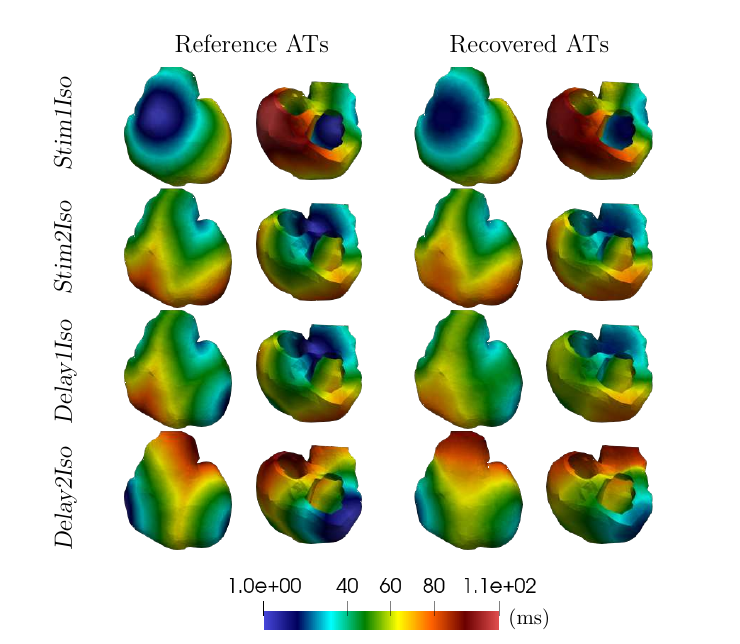}
    \caption{Reconstructed activation maps for the first four isotropic propagation test cases.}
    \label{fig:activation_iso}
\end{figure}

\begin{figure}[h!]
    \centering
    \includegraphics[width=0.7\linewidth]{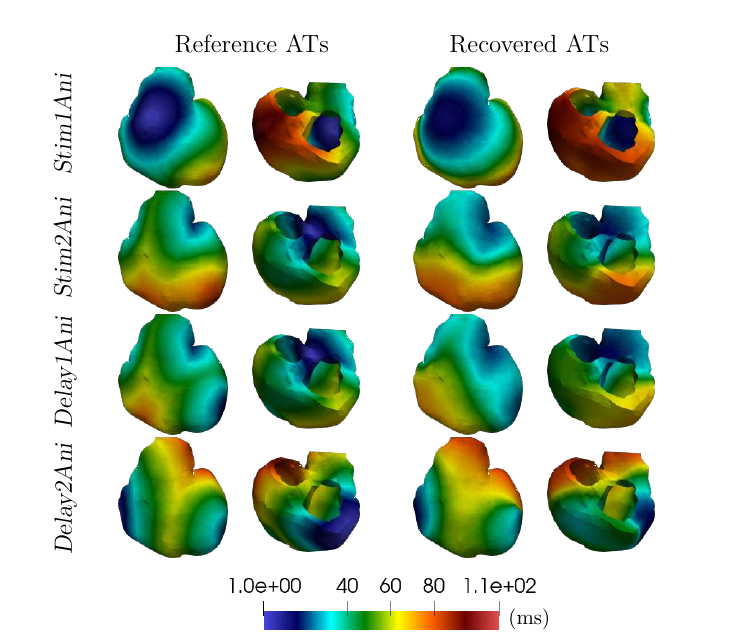}
	\caption{Reconstructed activation maps for the first four anisotropic propagation test cases.}\label{fig:activation_fibre}
\end{figure}

For the test cases \textsl{Stim1Iso}, \textsl{Stim2Iso}, \textsl{Delay1Iso} and \textsl{Delay2Iso}, the obtained activation maps are shown in Figure~\ref{fig:activation_iso} and for their anisotropic counterparts, the recovered activation maps are presented in Figure~\ref{fig:activation_fibre}. These visualizations offer an initial glimpse into the potential of the method. When cardiac conductivities are perfectly known in $O$ and in the distance $d$ (isotropic propagation test cases), the resulting reconstructed maps are of good quality, with a correlation coefficient with the reference greater than 0.97 across all four test cases. In the more realistic test case of inaccurate a priori conductivity tensors in the transfer matrix $O$ (i.e. anisotropic propagation test cases), the recovered activation maps are degraded, but key features, such as activation sites, remain clearly identifiable and the correlation coefficient remains above 0.95 for all cases. 

Nevertheless, we do not consider the reconstruction of the activation map to be the primary strength of the proposed particle filtering method, therefore we do not include a comparative study with more classical methods on this aspect. Our main objective is to develop nondeterministic tools. In the remainder of the paper, we focus on several probabilistic approaches that provide more informative insights than the raw activation maps.

\paragraph{Activation probability}

\begin{figure}
    \centering
    
    \begin{subfigure}{0.9\textwidth}
        \centering
        \includegraphics[width=\textwidth]{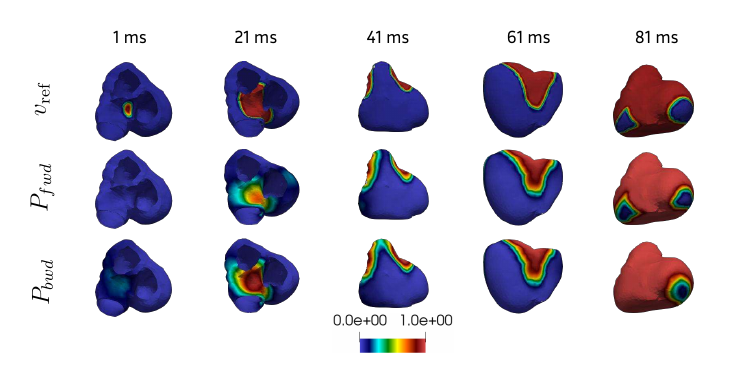}
        \caption{Case \textsl{Stim2Iso}}
        \label{fig:proba_activ_7}
    \end{subfigure}
    
    \vfill
    
    \begin{subfigure}{0.9\textwidth}
        \centering
        \includegraphics[width=\textwidth]{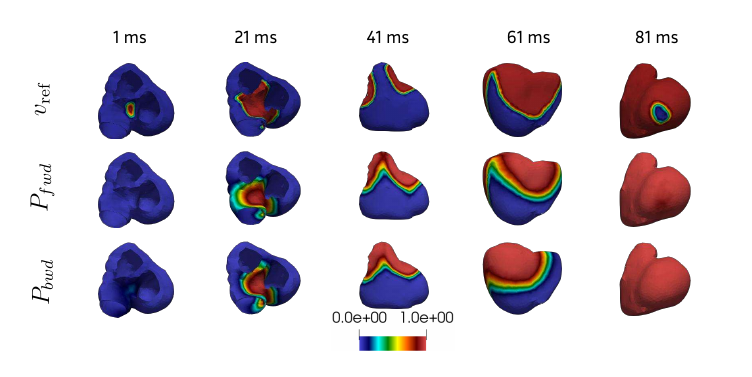}
        \caption{Case \textsl{Stim2Ani}}
        \label{fig:proba_activ_10}
    \end{subfigure}

    \caption{Probability of being activated at different times for stimulation \textsl{Stim 2}.}
    \label{fig:proab_activ_1}
\end{figure}

\begin{figure}
    \centering
    
    \begin{subfigure}{0.9\textwidth}
        \centering
        \includegraphics[width=\textwidth]{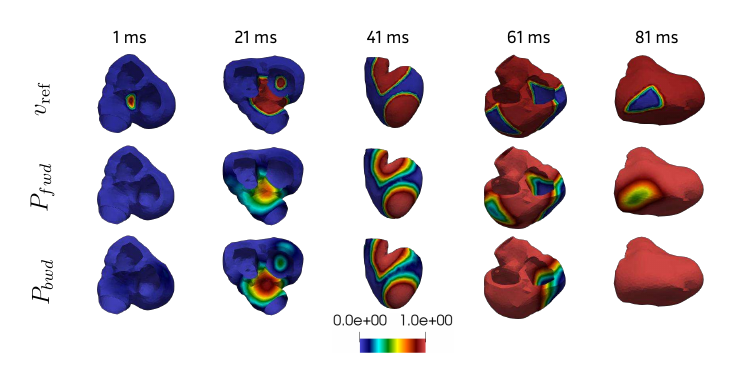}
        \caption{Case \textsl{Delay1Iso}}
        \label{fig:proba_activ_11}
    \end{subfigure}
    
    \vfill
    
    \begin{subfigure}{0.9\textwidth}
        \centering
        \includegraphics[width=\textwidth]{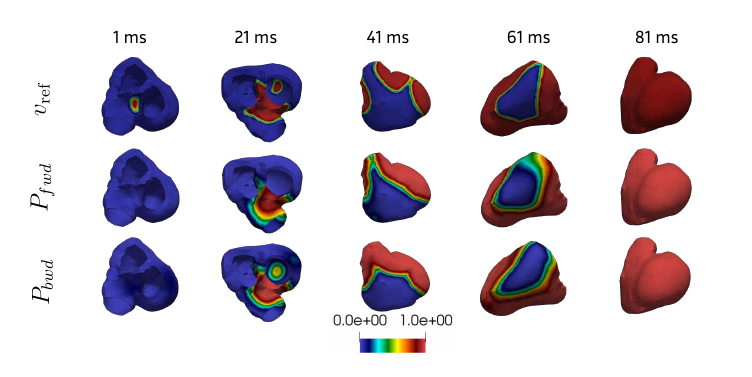}
		\caption{Case \textsl{Delay1Ani}}
		\label{fig:proba_activ_111}
    \end{subfigure}

	\caption{Probability of being activated at different times for stimulation  \textsl{Delay 1}.}\label{fig:proab_activ_2}
\end{figure}

\begin{figure}
    \centering
    
    \begin{subfigure}{0.9\textwidth}
        \centering
        \includegraphics[width=\textwidth]{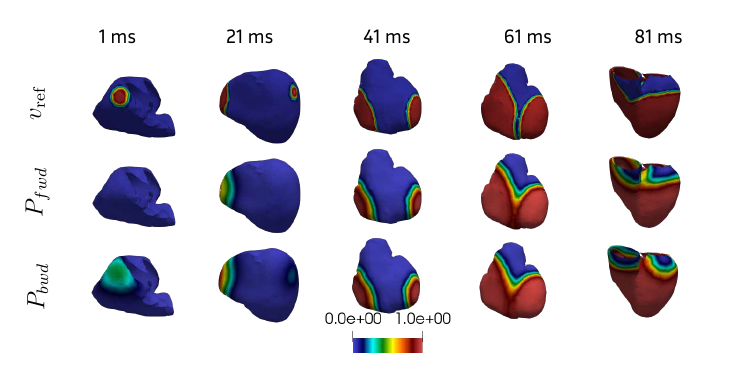}
        \caption{Case\textsl{Delay2Iso}}
        \label{fig:proba_activ_13}
    \end{subfigure}
    
    \vfill
    
    \begin{subfigure}{0.9\textwidth}
        \centering
        \includegraphics[width=\textwidth]{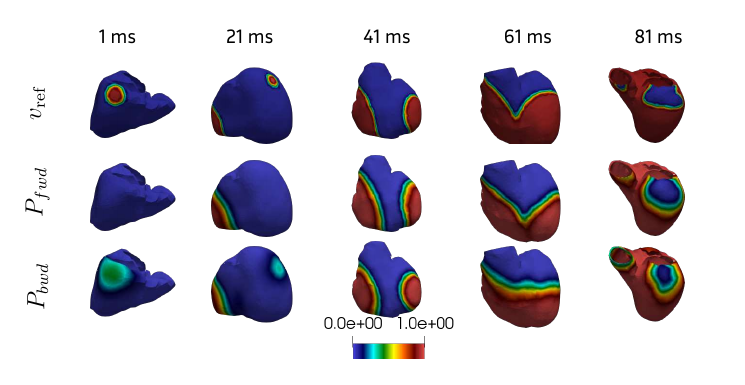}
		\caption{Case \textsl{Delay2Ani}}
		\label{fig:proba_activ_131}
    \end{subfigure}

	\caption{Probability of being activated at different times for stimulation  \textsl{Delay 2}.}\label{fig:proab_activ_3}
\end{figure}
Using the information provided by the particle filters, it is possible to evaluate, at each time step, the probability that each point of the mesh is activated. This is computed by summing, for each point, the weights of the particles in which the point is considered activated. A point is deemed activated if the transmembrane voltage in that point is superior to a threshold value of $0.5$. The probability of being activated at a point $x$ of the heart and at time $t_k$ is defined from the $N$ particles of the filter as 
\begin{equation}
	\mathbb{P}(v(x, t_k) > 0.5 | Y_{1:k}) \approx \sum\limits_{i = 1}^{N}{\omega^i_k \mathbbm{1}_{[0.5, 1]}(v_{\xi_k^i}(x))},
    \label{activation proba}
\end{equation} where $\mathbbm{1}_{D}$ is the indicator function of the domain $D$ included in $\mathbb{R}$.

Figures~\ref{fig:proab_activ_1}, \ref{fig:proab_activ_2}, and \ref{fig:proab_activ_3} present examples of such activation probability maps for various time steps and test cases. In each case, the activation probabilities computed using either the \textsl{forward} or the \textsl{backward} method are presented side by side with the true transmembrane voltage $v$, ranging between 0 and 1 due to the use of the Mitchell-Shaeffer ionic model.

Several interesting observations can be made. First, when the conductivity tensors in the heart are correctly known by the filter, we obtain fairly accurate results, as shown in Figure~\ref{fig:proba_activ_7}. However, as explained earlier, the \textsl{forward} method struggles to produce precise estimates in the early time steps, while the \textsl{backward} method is less accurate toward the end. It can then happen that the probability estimate is incorrect in some regions. For example, in Figure~\ref{fig:proba_activ_11}, the \textsl{forward} method misses an activation site at $21$ ms, and conversely, the \textsl{backward} method misses a resting area at $61$ ms. However, in these less accurate phases, the probability maps typically display regions with intermediate activation probabilities. Rather than being a limitation, this reflects the method’s capacity to represent and communicate uncertainty. Indeed, it is preferable to have a broad region of uncertainty that contains the true activation zone than to produce a single, deterministic yet misplaced solution.

This uncertainty becomes especially important in test cases where cardiac fibers are present in the true model but omitted in the filter's model. While such a modeling inaccuracy significantly degrades the deterministic activation maps, it has a less severe impact on the activation probability maps, which explicitly express confidence levels for each region. For instance, in Figure~\ref{fig:proba_activ_10}, the filter shows high confidence in the initial activation zone, which is accurately estimated, but lower certainty regarding the temporal progression of the activation front. 

The study of the \textsl{Delay} test cases, Figures \ref{fig:proab_activ_2} and \ref{fig:proab_activ_3},  provides additional insights. As expected, since the activations occur early in the sequence and are slightly staggered in time, the \textsl{forward} method struggles to correctly identify the activation sites. In contrast, the \textsl{backward} method proves more effective since both activation sites are already active at the final time step, and the filter can use subsequent observations to better infer the early stages. However, the \textsl{backward} method performs less well in estimating the final stages of the process. This asymmetry underscores the importance of having access to both the \textsl{forward} and \textsl{backward} outputs, which motivates the combined visualization tool presented in the following section.

\paragraph{Estimation of the earliest activation sites}
We now turn our attention to the estimation of the earliest activation sites, rather than the full activation sequence, with the same aim of providing a tool that incorporates confidence information. To the best of our knowledge, no such method has been reported in the literature. 

The center estimated by our method, as defined in~\eqref{eq:state_definition}, can be interpreted as a stand-in for these earliest activation sites. To quantify the likelihood of each mesh point to belong to an earliest activation site, we thus compute for each point a pseudo-probability $\tilde{P}(x)$ of being a center by summing the weights of all particles in which the point is selected as a center, for all time steps:
\begin{equation}
\begin{aligned}
	&\tilde{P}_k(x | Y_{1:k}) \overset{\mathcal{D}}{=}  \sum\limits_{i = 1}^{N}{\omega^i_k \sum\limits_{j = 1}^{l} \delta((c_k^j)^i = x)},\\
   & \tilde{P}(x) \overset{\mathcal{D}}{=} \sum\limits_{k = 1}^{n} \tilde{P}_k(x | Y_{1:k})
    \end{aligned}
\end{equation}where $(c_k^j)^i$ denotes the $j$-th center of the $i$-th particle at time $t_k$.

It is important to emphasize that this quantity is not a true probability, mostly because the values can exceed 1, as they range from 0 to the maximum number of centers represented in the state times the number of time steps. This is due to the fact that the actual number of distinct earliest activation sites is unknown, making it difficult to normalize these values into proper probabilities. Indeed multiple centers within a single particle may correspond to the same physical earliest activation site, and the number of times a center is present in a particle has no influence on the particle's likelihood.

\begin{figure}
    \centering
    
    \begin{subfigure}{0.49\textwidth}
        \centering
        \includegraphics[width=\textwidth]{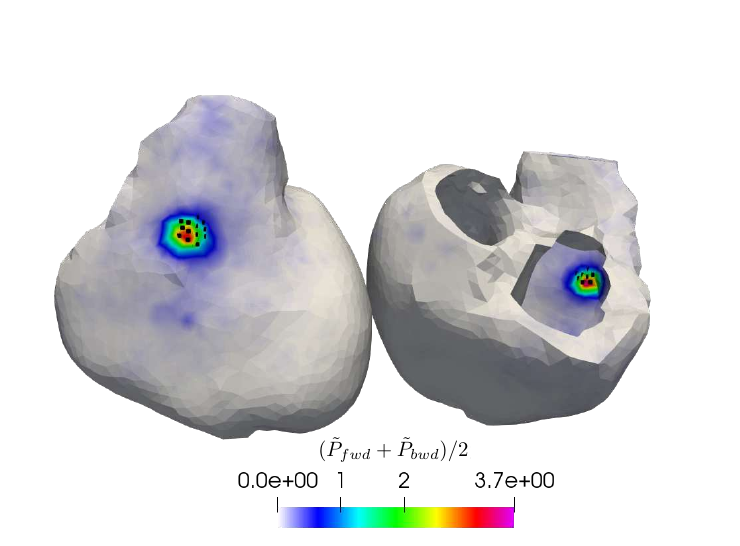}
        \caption{Case \textsl{Stim1Iso}}
        \label{fig:proba_center_6}
    \end{subfigure}
    \hfill
    \begin{subfigure}{0.49\textwidth}
        \centering
        \includegraphics[width=\textwidth]{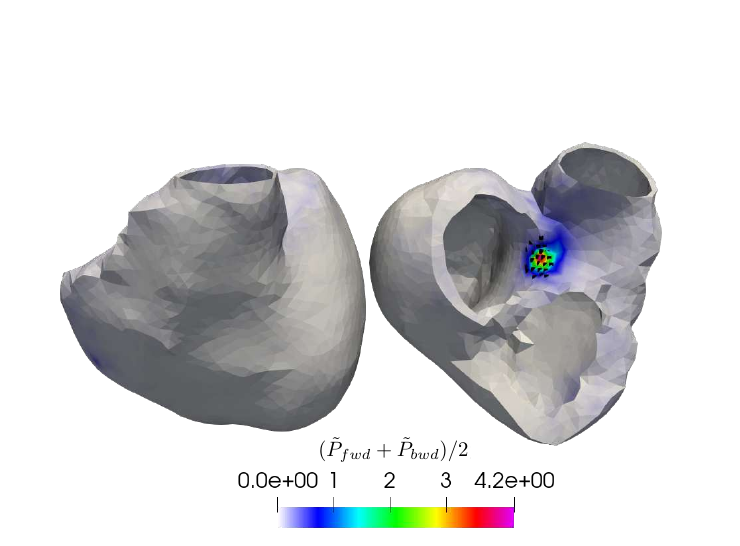}
        \caption{Case \textsl{Stim2Iso}}
        \label{fig:proba_center_7}
    \end{subfigure}
    
    \vfill
    
    \begin{subfigure}{0.49\textwidth}
        \centering
        \includegraphics[width=\textwidth]{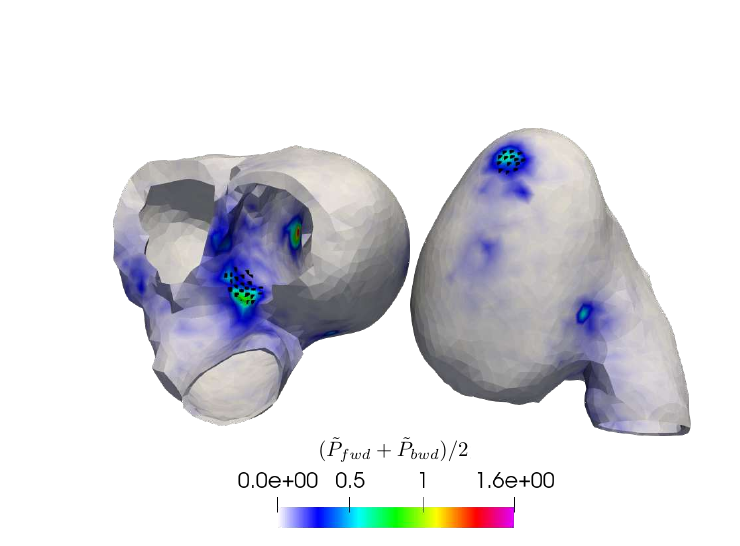}
        \caption{Case \textsl{Delay1Iso}}
        \label{fig:proba_center_11}
    \end{subfigure}
    \hfill
    \begin{subfigure}{0.49\textwidth}
        \centering
        \includegraphics[width=\textwidth]{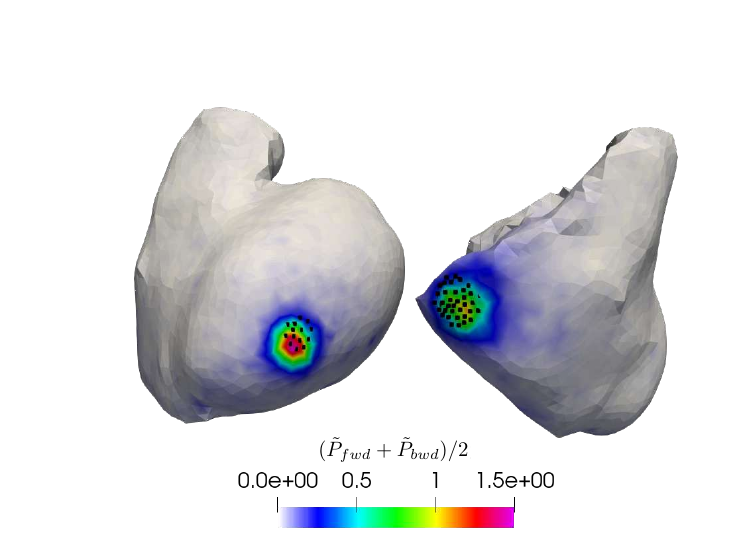}
        \caption{Case \textsl{Delay2Iso}}
        \label{fig:proba_center_13}
    \end{subfigure}
    
    \caption{Pseudo-probability of belonging to an initial activation site for the first four isotropic propagation test cases.}
    \label{fig:proba_center_iso}
\end{figure}

\begin{figure}
    \centering
    
    \begin{subfigure}{0.49\textwidth}
        \centering
        \includegraphics[width=\textwidth]{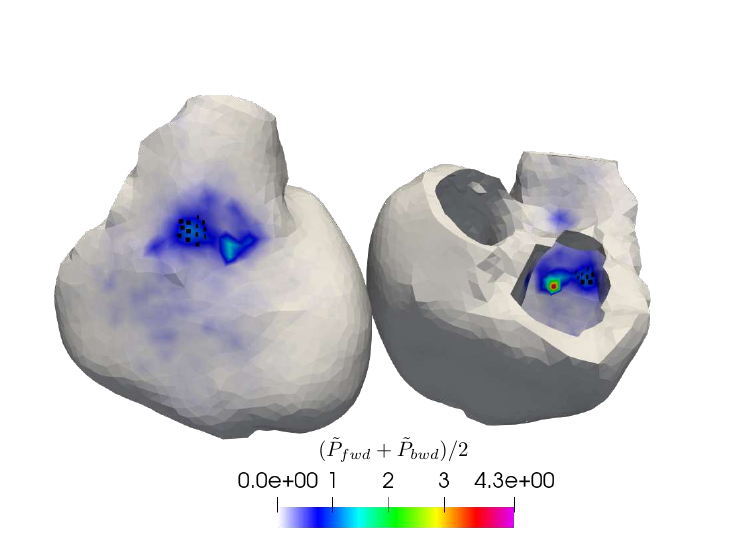}
        \caption{Case \textsl{Stim1Ani}}
        \label{fig:proba_center_9}
    \end{subfigure}
    \hfill
    \begin{subfigure}{0.49\textwidth}
        \centering
        \includegraphics[width=\textwidth]{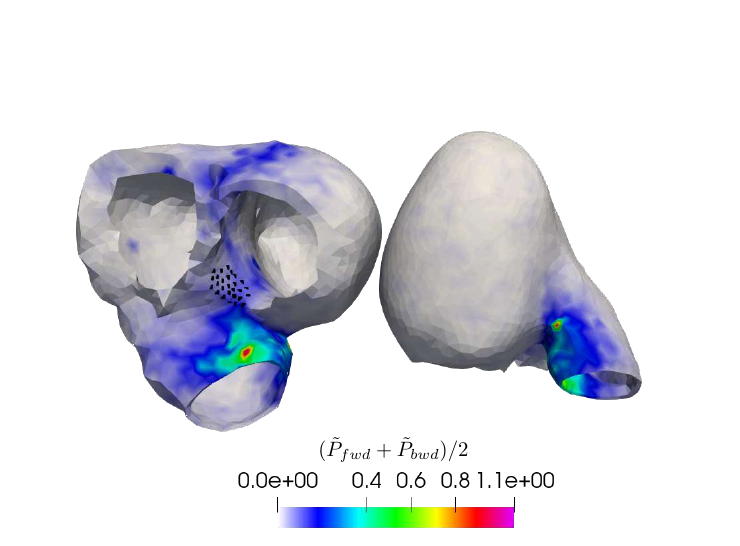}
        \caption{Case \textsl{Stim2Ani}}
        \label{fig:proba_center_10}
    \end{subfigure}
    
    \vfill
    
    \begin{subfigure}{0.49\textwidth}
        \centering
        \includegraphics[width=\textwidth]{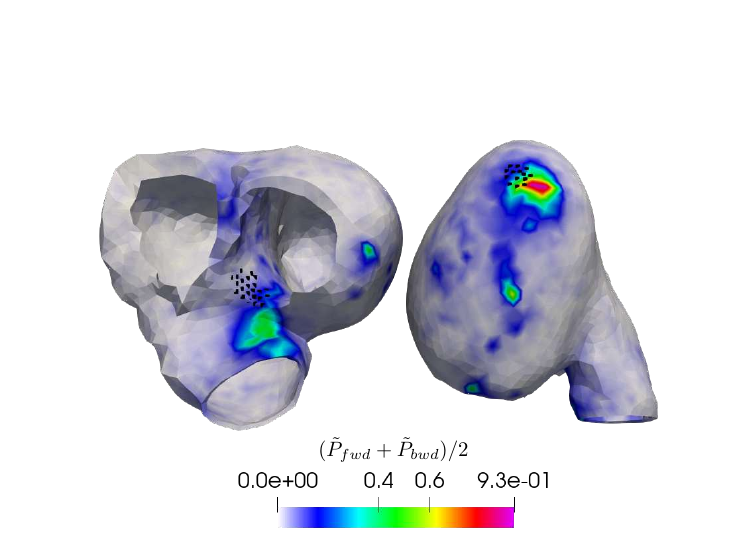}
        \caption{Case \textsl{Delay1Ani}}
        \label{fig:proba_center_111}
    \end{subfigure}
    \hfill
    \begin{subfigure}{0.49\textwidth}
        \centering
        \includegraphics[width=\textwidth]{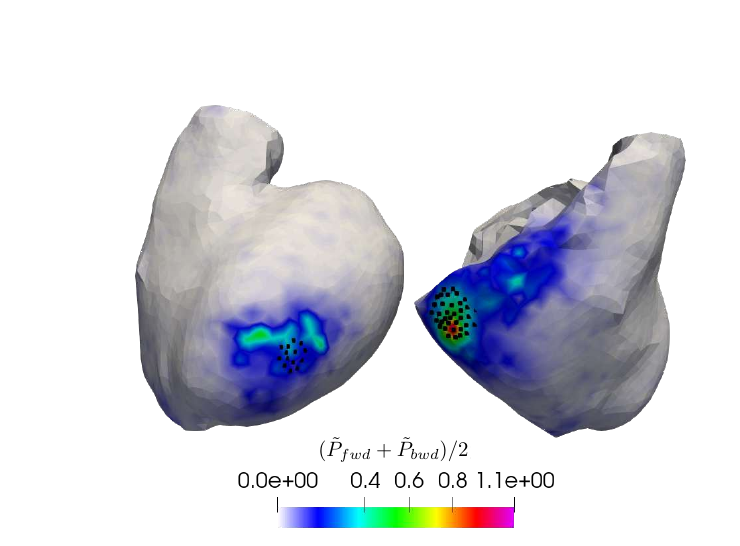}
        \caption{Case \textsl{Delay2Ani}}
        \label{fig:proba_center_131}
    \end{subfigure}
    
    \caption{Pseudo-probability of belonging to an initial activation site for the first four anisotropic propagation test cases.}
    \label{fig:proba_center_fibre}
\end{figure}

Figures~\ref{fig:proba_center_iso} and~\ref{fig:proba_center_fibre} show the maps of these pseudo-probabilities, obtained by averaging the results of the \textsl{forward} (index $fwd$) and \textsl{backward} (index $bwd$) methods, for the first four isotropic test cases and the first four anisotropic test cases respectively. On each map, the true stimulation sites are indicated by black dots. 

When the filter uses the exact electrical conductivity tensors (isotropic test cases, Figure~\ref{fig:proba_center_iso}), the earliest activation sites are located with high accuracy. In the test cases \textsl{Stim1Iso}, \textsl{Stim2Iso} and \textsl{Delay2Iso}, the true activation sites are even identified in a unique manner, meaning that no other possible stimulation sites appear on the map. In the \textsl{Delay1Iso} test case, where one of the two true stimulation sites is located near the right ventricular outflow tract, a hardly observable area of the heart, the two correct stimulation sites are identified among four possible sites.
 Near the right ventricular outflow tract, the potential stimulation zone appears larger, with a lower, more spread pseudo-probability. These results are  encouraging as we expect the filter to display all the likely possibilities, and not only one deterministic guess. Moreover, note that the temporal probability of being activated shown in Figure~\ref{fig:proba_activ_11} can also allow to discriminate true from false activation sites with a higher level of confidence.

In contrast, when significant modeling errors are introduced through the absence of fibers in the filter heart model (anisotropic data test cases, Figure~\ref{fig:proba_center_fibre}), the estimates are degraded, and the predicted centers tend to be slightly misplaced. However, this error is balanced by a lower confidence in the prediction and a correspondingly broader uncertainty region. Except in the cases where a stimulation occurs near the right ventricular outflow tract, where signal reconstruction proves particularly difficult (\textsl{Stim2Ani} and \textsl{Delay1Ani}), the true activation sites remain contained within the uncertainty region. This demonstrates the filter’s ability to effectively represent and communicate uncertainty in challenging settings.

In summary, the combined multiple visualization tools our filter offers may allow to refine our vision and understanding of the reconstructed output of ECGi.

\subsection{Distinguishing artificial from real conduction lines of block}\label{subsec:distinguish_block}

\begin{figure}[h!]
    \centering
    \includegraphics[width=0.7\linewidth]{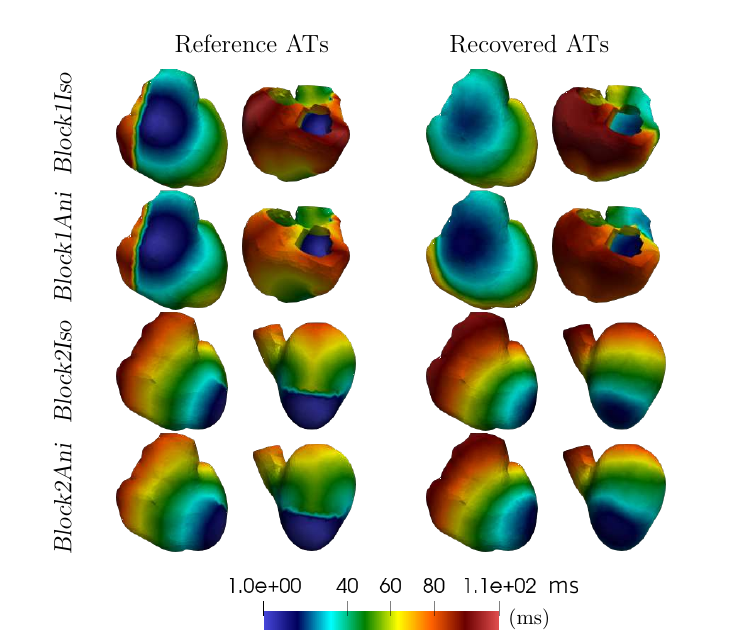}
    \caption{Activation maps obtained with only one homogeneous geodesic distance. The conduction lines of block cannot be reconstructed.}
    \label{fig:AT block bf}
\end{figure}

If we now look at the last $2 \times 2$ propagation test cases \textsl{Block 1} and \textsl{Block 2}, which include a real conduction line of block (see  details in Subsection~\ref{subsection:data_generation} and Table~\ref{tab:bidomain params}), we expect the filter not to be able to recover the lines of block in the activation maps. Indeed, our chosen parametrization of the activation front shape, trough balls in the isotropic and homogeneous geodesic distance, prevents from recovering a non circular activation shape in the myocardium. This is clearly observed in the recovered activation maps presented in Figure~\ref{fig:AT block bf} for the \textsl{Block1Iso}, \textsl{Block1Ani}, \textsl{Block2Iso} and \textsl{Block2Ani} test cases. 

Nevertheless, recovering true regions of low conduction velocity is of interest to clinicians as it can indicate pathology. Furthermore, it is well known that the classical Cauchy-Tikhonov-based approach to ECGi tends to produce too many lines of block in the reconstructed activation maps, among which some are true, and many are artifacts \cite{schuler2021reducing, duchateau2019performance}. It would thus be interesting to overcome the intrinsic limitation of our particle filtering model, while offering a tool for discriminating true from artificial conduction lines of block in the activation maps.

We thus propose to adapt our method, not to detect lines of block directly, but to help distinguish between artificial and true lines of block. Assuming that a pre-existing activation map has been computed using any method (e.g. solving the Cauchy problem for the Laplace equation), and that lines of block have been identified on this map, it is possible to compute new geodesic distances and operator $O$ that account for the underlying changes in conductivity. In that case, electrical conductivities in \eqref{sigma_distance} and in $O$ can be tuned to produce a distance map containing a line of block similar, though not perfectly identical, to the one in the activation map.

We then adapt the filtering method by extending the state defined in \eqref{eq:state_definition} with a discrete component. This discrete state allows each particle to choose from a predefined set of geodesic distances and corresponding transfer matrix $O$. The distribution of particles across the different geodesic metrics, those with or without a line of block for instance, provides an evaluation of the confidence in the actual presence of such a conduction block in the heart. Denoting by  $m_k^i$ the mode, composed of a geodesic distance and its associated transfer matrix, of particle $i$ at time \( t_k \), the probability reconstructed by the filter that a given geodesic distance \( d \) is valid at time \( t_k \) is given by:
\begin{equation}
    p_k(d) = \sum_{i=1}^N \omega_k^i \, \delta_{m_k^i}(d).\label{eq:mode_proba}
\end{equation}

In practice, each particle is initialized by uniformly selecting one of the available geodesic distances. Then, at each time step, each particle has a 99\% chance of keeping the same distance. Otherwise, it uniformly selects a new distance from the available options. Allowing particles to switch modes not only enables them to correct themselves over time, but also allows the filter to test the presence of multiple candidate lines of block, even if no single geodesic metric accounts for all of them simultaneously. Indeed, the filter's estimate can switch between distances based on which one best fits the data at each time step. In addition to this mode evolution, the resampling steps tend to duplicate particles associated with the most probable distance metric and corresponding operator~$O$.

Using approximations of the geodesic distances based on the \textsl{Block 1} and \textsl{Block 2} test cases, we ran 50 \textsl{forward} and 50 \textsl{backward} filters on the 2$\times$2 test cases that include a line of block. Since each run is independent, the computation can be fully parallelized, avoiding any additional computation time.  Each filter was provided with the two distance maps derived from the \textsl{Block 1} and \textsl{Block 2} test cases, as well as the isotropic distance. We thus expect the filter to be able to have a high confidence in the presence of the true line of block, but also to discard the false one. Figure~\ref{fig:courbes blocs} shows, for each test case, the mode distribution, computed using \eqref{eq:mode_proba}, averaged across all filters. As for earliest activation estimate, computing an average across the \textsl{forward} and \textsl{backward} method allows us to combine the strengths of those methods.

\begin{figure}[htbp]
  \centering
  \begin{tikzpicture}
    \begin{axis}[
      hide axis,
      xmin=0, xmax=1,
      ymin=0, ymax=1,
      legend columns=3,
      legend style={
        draw=none,
        font=\footnotesize,
        /tikz/every even column/.append style={column sep=1em},
        at={(0.5,1.0)},
        anchor=south
      }
    ]
      \addlegendimage{blue, thick}\addlegendentry{True line of block}
      \addlegendimage{red, thick}\addlegendentry{False line of block}
      \addlegendimage{green!60!black, thick}\addlegendentry{No line of block}
    \end{axis}
  \end{tikzpicture}

  \vspace{0.5em}  
  
  \begin{subfigure}[t]{0.51\textwidth}
    \centering
    \myplotTL
    \caption{\textsl{Block1Iso}}\label{fig:proba_Block1Iso}
  \end{subfigure}
    \hspace{-0.04\textwidth}  
    \begin{subfigure}[t]{0.51\textwidth}
    \centering
    \myplotTR
    \caption{\textsl{Block2Iso}}\label{fig:proba_Block2Iso}
  \end{subfigure}

  \vspace{1em}

  \begin{subfigure}[t]{0.51\textwidth}
    \centering
    \myplotBL
    \caption{\textsl{Block1Ani}}\label{fig:proba_Block1Ani}
    \label{41}
  \end{subfigure}
  \hspace{-0.04\textwidth}
  \begin{subfigure}[t]{0.51\textwidth}
    \centering
    \myplotBR
    \caption{\textsl{Block2Ani}}\label{fig:proba_Block2Ani}
  \end{subfigure}

\caption{Evolution of the probability distribution over geodesic distance modes, averaged across 50 \textsl{forward} and 50 \textsl{backward} filter outputs, for test cases containing a true block line. The blue line indicates the actual block line present in the heart for each case.}
  \label{fig:courbes blocs}
\end{figure}

For test cases based on \textsl{Block 2}, regardless of the modeling error due to the presence of fibers, the filters show a very high confidence in the presence of the real line of block, as illustrated in Figures~\ref{fig:proba_Block1Iso} and~\ref{fig:proba_Block2Iso}. In contrast, the block from \textsl{Block 1} test cases is more difficult to detect with high confidence. For filters that do not make an assumption error regarding the presence of fibers, the filter is confident in the presence of the real block during the first $6$0~ms, as shown in Figure~\ref{fig:proba_Block1Iso}. Beyond this point, all geodesic distances exhibit roughly the same likelihood. According to the activation maps in Figure~\ref{fig:AT block bf}, the $60$~ms mark corresponds to the moment when the block is crossed, after which its effect becomes very weak. As the confidence level in the first part is lower than that of the filter associated with the \textsl{Block 2} test cases, this explains the indecision of the filter in the latter part of the simulation.
For test case \textsl{Block1Ani}, due to the lower observability of the block combined with the modeling error caused by the omission of fibers, the filters tend to overlook the real block and instead show confidence in the absence of a block, as illustrated in Figure~\ref{fig:proba_Block1Ani}. 
Note, however, that throughout the various test cases, the filters have consistently been able to discard the false block, regardless of the modeling errors.

\begin{figure}[h!]
    \centering
    \includegraphics[width=0.7\linewidth]{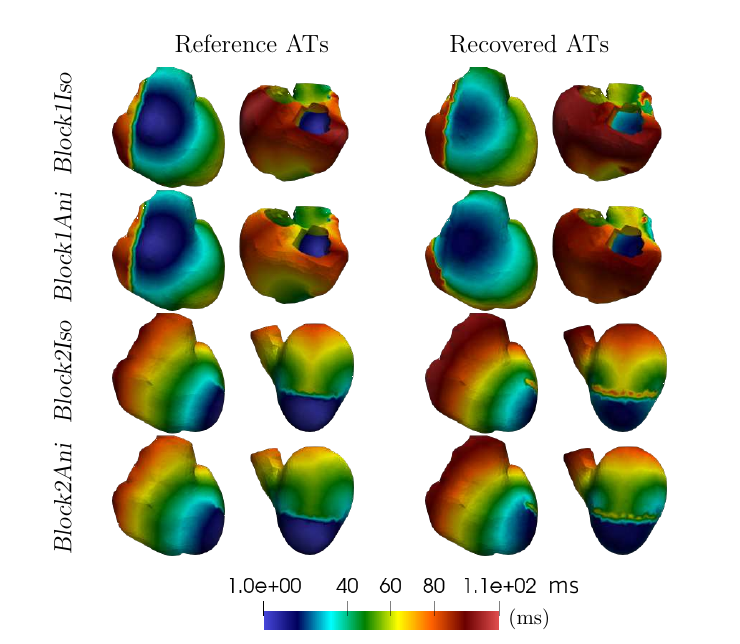}
    \caption{Activation maps obtained with an extended state which includes the choice between three geodesic distances: one homogeneous, one heterogeneous containing the true a priori conduction line of block and one heterogeneous containing a false a priori conduction line of block. The true conduction lines of block are recovered in the maps or the four test cases.}
    \label{fig:AT block after}
\end{figure}
\begin{figure}[h!]
    \centering
    \includegraphics[width=0.35\linewidth]{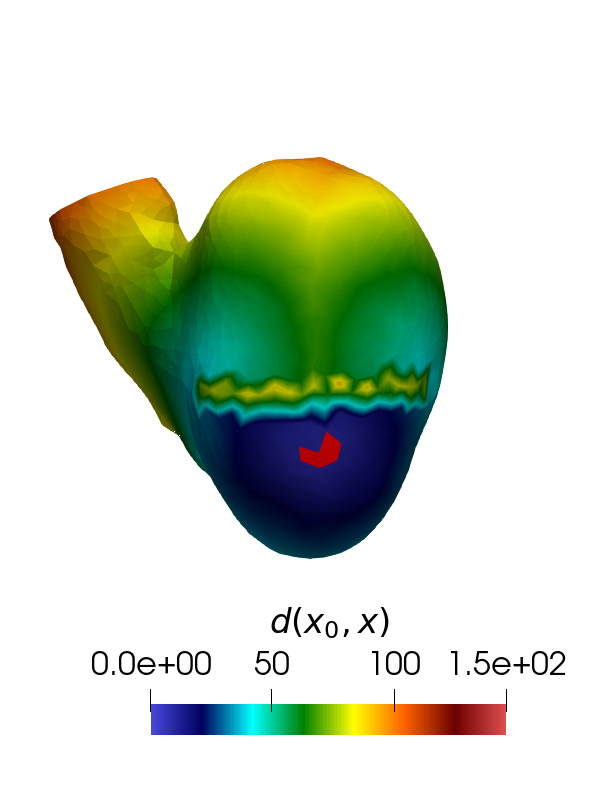}
    \caption{Example of the computed geodesic distance from the points in the red zone, with an a priori block. The reconstruction method of equation \eqref{sigma_distance}, intentionally imperfect, produces the block’s atypical shape.}
    \label{fig:ex_dist_block}
\end{figure}

Activation maps can also be recomputed, using the output of the extended-state filter. Such maps are shown in Figure~\ref{fig:AT block after}. With the extended state and the choice between three possible distance maps and operators $O$, we are able to reconstruct accurate activation maps, containing the true conduction lines of block for all four test cases \textsl{Block1Iso}, \textsl{Block1Ani}, \textsl{Block2Iso} and \textsl{Block2Ani}. For the  \textsl{Block1Ani} test case, whose mode distribution across time did not indicate the presence of a true line of block (see Figure~\ref{41}), we are still able to observe a reconstructed line of block in the activation map, though smaller than in the reference map. For the \textsl{Block2Iso} and \textsl{Block2Ani} cases, the line of block localization and presence is well recovered. Some singularities appear on both sides of the discontinuity line. These are in fact caused by approximations in the reconstructed geodesic distance, which, for instance, may exaggerate the strength of the block line compared to the real one. Figure~\ref{fig:ex_dist_block} provides a glimpse of these imperfections in the geodesic distance computed using formula~\eqref{sigma_distance}. It shows the resulting geodesic distance from a specific region  to all points of the heart surface, reconstructed to mimic \textit{Block 1}. While the presence of the conduction line of block is clearly visible, it appears wider and stronger than in the reference activation map of Figure~\ref{fig:AT block after}. In addition, its shape is irregular and dependent on the underlying mesh discretization, thereby explaining the artificial geometrical features observed in the reconstructed activation maps.

Overall, these results tend to validate the ability of the extended filter to discriminate true from artificial lines of block, even with imprecision in the estimated distance matrix and conductivity tensors in $O$.

\section{Conclusion}

In this paper, we introduced a new method for estimating various characteristics of the cardiac activation sequence from body surface potential maps. This approach is based on particle filtering techniques, which offer several advantages. First, it allows for more complex transition and observation models than the standard linear Gaussian assumptions, making it possible to incorporate more realistic noise distributions. Second, it avoids the need to assume a predefined shape for the posterior distribution, as is required by Kalman-based filters. This added flexibility enables a more nuanced evaluation of the confidence in the estimates, beyond a simple covariance matrix. 

In addition to the classical activation map, we proposed three new estimators based on probabilistic interpretations. The first provides, for each point and time step, the probability of being activated at that time. This estimator offers a detailed view of the activation process, enabling the identification of activation sites and activation times, along with an assessment of the confidence in these estimates. To offer a simpler tool for identifying the earliest activation sites and evaluating their reliability, we introduced a map that assigns to each point a pseudo-probability of belonging to an earliest activation site. Finally, to complement existing deterministic approaches, we proposed a method for assessing whether a detected line of block is likely to be physiological or an artifact, thereby introducing a probabilistic dimension to otherwise deterministic frameworks.

We tested our method on several simulated test cases using a myocardial volume mesh, though the method could equally be applied to a surface mesh of the heart.
In the most optimistic cases, where the conductivities of the heart tissue were known, the method was able to recover the correct number of earliest activation sites, through the temporal activation probability maps (Figures~\ref{fig:proab_activ_1},~\ref{fig:proab_activ_2},~\ref{fig:proab_activ_3}). At the same time, the true earliest activation sites lay within high pseudo-probability regions of the map indicating the likelihood of belonging to an earliest activation site (Figure~\ref{fig:proba_center_iso}). In fact, rather than being opposed, these different views can be combined to improve the interpretation of the results, for example by discarding incoherent earliest activation sites.
To better mimic real-world situations, where these conductivities are typically unknown, we also considered cases with fibers present in the heart during data generation but not accounted for in the filter. Even with the introduction of this model mismatch, our method still produced satisfactory results. Although the estimates were often shifted, they generally remained within the computed uncertainty regions, whether for the activation time at a given point or for the earliest activation sites. Moreover, the method proved effective in estimating the number of earliest activation sites.

Due to the imposed shape of the activation front, made of overlapping activation balls, our method still faces some limitations. For example, if a line of block is present in the heart but not modeled in the filter, the method fails to detect it. To address this issue, we proposed an extension that successfully overcomes the problem. Other challenging issues will likely arise in practical applications, but similar state extensions could be employed to address them effectively.

Though we voluntarily introduced modeling errors in the filter model, we still used simulated data, which are well known to produce ECGi reconstructions of higher quality than clinical or experimental data. Indeed, in a clinical setting, a lot of unknowns, modeling errors and noise, add to the difficulty of the ECGi problem. Moreover, real-word data are usually obtained through $148$ to $255$ electrodes vests, which is way less than the approximately $1000$ body surface data points of our torso surface mesh. Our methodology clearly extends to the usual torso electrode vest, provided a coarse mesh of the torso is available, but needs to be assessed in such a setup. Thus future work should focus on evaluating our new method on wider and more challenging experimental datasets. As with most methodological developments in ECGi, a practical adaptation to clinical data would require reintroducing biophysical dimensions into the prescribed transmembrane voltage $V$, as well as developing an appropriate mesh-processing pipeline. Starting from patient-specific heart and torso geometries together with body-surface electrode positions, it would be necessary to generate a complete heart–torso mesh in which the data locations coincide with nodes of the torso surface mesh. Moreover, for the method to effectively distinguish true lines of block from artifacts, an additional step would be required to automatically derive geodesic distances that account for conduction blocks from a given activation map.

Finally, it may be worth considering alternative particle filtering schemes, potentially more efficient than the basic SIR filter. It should also be noted that certain a priori information can be added to the filter in order to improve its performance. One could imagine introducing a priori fiber directions into the transfer matrix model $O$ and into the geodesic distances, or using a prior deterministic ECGi reconstruction to deduce the initialization state $X_0$. 

\section*{Acknowledgments}

Experiments presented in this paper were carried out using the
PLAFRIM experimental testbed, being developed under the Inria PlaFRIM development
action with support from Bordeaux INP, LABRI and IMB and other entities: Conseil
Régional d’Aquitaine, Université de Bordeaux and CNRS (and ANR in accordance to the
programme d’investissements d’Avenir (see http://www.plafrim.fr/).

The authors would like to warmly thank Lisl Weynans and Yves Coudière for taking the time to proofread the manuscript.

\bibliographystyle{unsrt}  
\bibliography{main}      

\end{document}